\documentclass[fleqn,usenatbib]{mnras}

\usepackage{newtxtext,newtxmath}

\usepackage[T1]{fontenc}

\DeclareRobustCommand{\VAN}[3]{#2}
\let\VANthebibliography\thebibliography
\def\thebibliography{\DeclareRobustCommand{\VAN}[3]{##3}\VANthebibliography}

\usepackage{graphicx}	
\usepackage{amsmath}	

\title[$\mathrm{M_{BH}}$--$\sigma$ vs.\ $\mathrm{M_{BH}}$--$\mathrm{M_{bulge}}$]{The Impact of Black Hole Scaling Relation Assumptions on the Mass Density of Black Holes}

\author[C. Matt et al.]{
Cayenne Matt,$^{1}$\thanks{E-mail: cayenne@umich.edu}
Kayhan G{\"u}ltekin,$^{1}$
Joseph Simon$^{2}$\thanks{NSF Astronomy and Astrophysics Postdoctoral Fellow}\\
$^{1}$Department of Astronomy, University of Michigan, 1085 S. University Ann Arbor, MI 48109, USA
\\
$^{2}$Department of Astrophysical and Planetary Sciences, University of Colorado, Boulder, CO 80309, USA
}

\date{Accepted XXX. Received YYY; in original form ZZZ}

\pubyear{2023}

\newcommand{\msigma}{$\mathrm{M_{BH}}$--$\sigma$}

\newcommand{\mmb}{$\mathrm{M_{BH}}$--$\mathrm{M_{bulge}}$}

\newcommand{\rev}{}

\begin{document}
\label{firstpage}
\pagerange{\pageref{firstpage}--\pageref{lastpage}}
\maketitle

\begin{abstract}
We examine the effect of supermassive black hole (SMBH) mass scaling relation choice on the inferred SMBH mass population since redshift $z \sim 3$. To make robust predictions for the gravitational wave background (GWB) we must have a solid understanding of the underlying SMBH demographics. Using the SDSS and 3D HST+CANDELS surveys for $0 < z < 3$ we evaluate the inferred SMBH masses from two SMBH-galaxy scaling relations: \mmb\ and \msigma.  Our SMBH mass functions come directly from stellar mass measurements for \mmb, and indirectly from stellar mass and galaxy radius measurements along with the galaxy mass fundamental plane for \msigma.
We find that there is a substantial difference in predictions especially for $z > 1$, and this difference increases out to $z = 3$. In particular we find that using velocity dispersion predicts a greater number of SMBHs with masses greater than $10^9 \mathrm{M}_\odot$. \rev{The GWB that pulsar timing arrays find evidence for is higher in amplitude than expected from GWB predictions which rely on high redshift extrapolations of local SMBH mass-galaxy scaling relations.} The difference in SMBH demographics resulting from different scaling relations may be the origin for the mismatch between the signal amplitude and predictions. Generally, our results suggest that a deeper understanding of the potential redshift evolution of these relations is needed if we are to draw significant insight from their predictions at $z > 1$. 

\end{abstract}




\section{Introduction}
Supermassive black holes (SMBHs) reside in the nuclei of nearly all massive galaxies \citep[see, e.g.,][]{Kormendy_Ho_2013}. Through galaxy mergers, these SMBHs can form dual and binary SMBHs \citep{Begelman_1980}. In the final stages of their evolution, before coalescence, SMBH binaries lose energy and angular momentum purely though gravitational waves (GW). The combined \rev{GW} signal from SMBH binaries is expected to be a stochastic background known as the gravitational wave background \citep[GWB;][]{Press_1972, Sesana_2004, Burke_Spolaor_2019}. Though \rev{GW} detectors such as LIGO, VIRGO, and KAGRA have successfully detected many GW events from stellar mass compact objects \citep{LIGO_2015, VIRGO_2015, KAGRA_2021}, the frequency range of GWs emitted by SMBH binaries is far below even the lowest detectable limit for Earth-based detectors. For such GWs, a much longer baseline is needed. \rev{To achieve this, pulsar timing arrays \citep[PTA; e.g.,][]{Sazhin_1978, Detweiler_1979, Foster_1990} use high-precision time-of-arrival measurements of millisecond pulsars to measure the change in Earth--pulsar distances for $\sim\mathrm{kpc}$-scale baselines. There are several years-long PTA campaigns, including North American Nanohertz Observatory for Gravitational Waves \citep[NANOGrav;][]{Ransom_2019}, European Pulsar Timing Array \citep{Perera_2017}, Parkes Pulsar Timing Array \citep{Goncharov_2021}, and Chinese Pulsar Timing Array \citep{Lee_2016}, Indian Pulsar Timing Array \citep{Nobelson_2022}, South Africa Pulsar Timing Array \citep{Spiewak_2022}.}

\rev{Several PTAs have individually  made significant progress towards detecting the GWB with evidence for a GWB with the characteristic quadrupolar signal of GWs \citep{Hellings_1983, Agazie_2023, Antoniadis_2023, Reardon_2023, Xu_2023}. Previously, the NANOGrav 12.5-year data \citep{Arzoumanian_2020}, while not having sufficient signal-to-noise to see the \citep{Hellings_1983} correlation, showed a common red noise process that shared many traits characteristic of the expected GWB. NANOGrav's signal, however, is significantly higher in amplitude than many predictions of the GWB \citep{Arzoumanian_2020, Shannon_2015, McWilliams_2014, Middleton_2021, Zhu_2019, Bonetti_2018, Agazie_2023}. The newest PTA data increase the significance of the high-amplitude GWB with support for characteristic strain amplitude of $h_c \sim 2 \times 10^{-15}$ consistent in all of the data sets finding evidence for \citep{Hellings_1983} correlations \citep{Agazie_2023, Antoniadis_2023, Reardon_2023, Xu_2023}.  In fact, three of the analyses are inconsistent with $h_c \le 1 \times 10^{-15}$ \citep{Agazie_2023, Antoniadis_2023, Reardon_2023}. The discrepancy between high amplitude observed and that expected from SMBH binaries has been explained with exotic theories such as cosmic strings \citep{Infante_2000, Ellis_2021} and inflationary universe models \citep{, Vagnozzi_2021, Allen_1988},  or extreme parameterizations of our current models \citep{Middleton_2021}. This opens the possibility that the explanations for the GWB signal should be revised \citep{Simon_2023, AgazieBHB_2023}.}

\rev{Though there are many SMBH properties that influence the emitted GWs, the mass distribution of SMBHs is fundamentally linked to the characteristic strain amplitude of the GWB and may be the most significant contributor to the amplitude we observe. \citet{Phinney_2001} noted that the characteristic strain amplitude from an isotropic background of binary SMBHs depends on four key quantities:
(i) the chirp mass of the binary, $\mathcal{M}^{5 / 3} \equiv \mathrm{M}_1 \mathrm{M}_2\left(\mathrm{M}_1+\mathrm{M}_2\right)^{-1 / 3}$, where $\mathrm{M}_1, \mathrm{M}_2$ are the masses of the SMBHs in the system with $\mathrm{M}_1 \geq \mathrm{M}_2$; (ii) the frequency of the emitted \rev{GWs}, $f$, which is twice the orbital frequency; (iii) the present-day comoving number density of merged remnants, $N_0$; and (iv) the redshift, $z$ as
\begin{equation}
    h_c \sim \mathcal{M}^{5/6} f^{-2/3} N_0^{1/2} \langle(1+z)^{-1 / 6}\rangle.
    \label{hc}
\end{equation}
Note that the amplitude has the strongest dependence on chirp mass, and so the signal is dominated by the most massive black holes. Below $z = 1$ the PTA band is dominated by local SMBH binaries, but the GWB amplitude is additionally influenced by \rev{galaxies} that merged at higher redshifts. SMBH evolution is determined, among other things, by mass and so a higher mass population of SMBHs at $z > 1$ may reflect a higher redshift evolution, thus the astrophysical history of SMBH mass evolution is encoded in the GWB.}

Since direct measurements of SMBH masses are only possible for nearby sources, we are often left to infer masses from properties of their host galaxies \citep{Richstone_1998}. There exist\rev{s} a wealth of relations between galaxy properties and the mass of their central black hole, all with varying degrees of scatter \citep[discussed further in ][]{Kormendy_Ho_2013}. Here we focus on two relations in particular: the correlation between SMBH mass with velocity dispersion ($\sigma$) and bulge stellar mass (M$_\mathrm{bulge}$). In the local universe, despite \msigma\ having lower scatter \citep{Gebhardt_2000, Gebhardt_2003, Ferrarese_2000, Kormendy_Ho_2013, Mcconnell_Ma_2013, Gultekin_2009}, both relations were found to be remarkably accurate when reproducing known SMBH masses from either stellar mass or velocity dispersion. \rev{These scaling relations are based on direct, dynamical mass measurements, which have been shown to be robust. For example, SMBH mass estimates in M87 have previously had discrepancies up to a factor of 2.5 when using stellar kinematics \citep{Gebhardt_2011} versus gas dynamics \citep{Ford_1994, Walsh_2013}. These are now seen as due to gas filaments \citep{Osorno_2023} which agrees with the mass found by the Event Horizon Telescope collaboration \citep{EHT_2019}.}

While there is general agreement in the local universe between SMBH masses predicted from stellar mass and velocity dispersion, it is worth discussing instances where these relations are thought to break down. Though we do not investigate it in this paper, SMBH mass is well-predicted from host luminosity. When investigating SMBH masses of large, luminous, brightest cluster galaxies (BCGs), \citet{Lauer_2007_sigma} found that \msigma\ fails to reproduce the extreme masses above $\mathrm{M}_\mathrm{BH} \sim 3 \times 10^9 \mathrm{M}_\odot$ measured and predicted from M$_\text{BH}$--$L$. Similarly, \citet{Mcconnell_Ma_2013} discuss this same trend, which they call a ``saturation'' effect, for which not only \msigma, but also \mmb\  under-predict the highest mass SMBHs in core galaxies. Both relations display this saturation at the high end of the relations that is not seen in the  M$_\text{BH}$--$L$.

We see a strikingly different pattern, however, when considering red nugget galaxies---galaxies with relatively small radii for their masses and high velocity dispersions that are more typical of younger galaxies. 
Red nugget galaxies may be representative of the high-redshift galaxy population, possibly because they have avoided mergers for a large portion of their lives \citep{Quilis_2013}. One red nugget is NGC 1277 which hosts a SMBH with a mass of $(4.9 \pm 1.6) \times 10^9 \mathrm{M}_\odot$ \citep{Walsh_2016}. NGC 1277's SMBH is over massive compared to the total stellar mass of the galaxy ($1.2 \times 10^{11} \mathrm{M}_\odot$) and is an outlier in the M$_\text{BH}$--$\text{M}_\mathrm{bulge}$ relation which predicts a mass of around $(4.9$--$6.23) \times 10^8 \mathrm{M}_\odot$. However, because of its high velocity dispersion, \msigma\ reproduces the measured SMBH mass more accurately, predicting a mass of $(2.9$--$3.7) \times 10^9 \mathrm{M}_\odot$ and the dynamical mass lies within the intrinsic scatter of the relation \citep{Kormendy_Ho_2013,Kormendy_Bender_2013,Forrest_2022,van_den_Bosch_2012}. Recently, it has been found that NGC 1277 may have lost the majority of its dark matter, suggesting an alternative evolutionary path \citep{Comeron_2023}, but NGC 1277 is not the only galaxy for which $\sigma$ has been found to be a better predictor of SMBH mass. MRK 1216  is another one of several well studied examples of this type of object which exhibit similar traits \citep{Y_ld_r_m_2015, Ferre_Mateu_2015, Ferre_Mateu_2017}.

Despite the great promise of the \msigma\ relation as a SMBH mass predictor, it is resource intensive to measure velocity dispersion at high redshift due the spectral quality required to resolve the necessary spectral features. To overcome this, the \mmb\  method is commonly used because it relates the relatively easily measured bulge stellar mass directly to the SMBH mass. \rev{This relationship \rev{is well measured} within our local universe, but a more accurate mass predictor may be needed for high redshifts ($z > 1$) where the a significant fraction of the GWB signal originates}.

To circumvent the spectral limitations on measuring velocity dispersion, in this paper we use the mass fundamental plane (MFP) of galaxies, which links total stellar mass and half light radius to stellar velocity dispersion. The MFP therefore allows us to infer velocity dispersion for distant galaxies and thus extend the \msigma\ relationship to higher redshifts. \citet{van_der_Wel_2014} investigated the evolution of the relationship between galaxy total stellar mass ($\mathrm{M}_*$) and effective radius ($\mathrm{R}_\mathrm{eff}$). They found that galaxy masses do not evolve along the $z = 0$ $\mathrm{M}_*$--$\mathrm{R}_\mathrm{eff}$ relation, but from redshift 0 to 3, the effective radii decrease substantially. This evolution of the $\mathrm{M}_*$--$\mathrm{R}_\mathrm{eff}$ relationship indicates that galaxies start off relatively compact and become more diffuse as they age as a result of mergers, feedback processes, and other galaxy interactions.  This change in radius is not incorporated in any way into the \mmb\  relation. Applying the local \mmb\ relationship to high-redshift galaxies results in a relatively unchanging SMBH mass population throughout time.

Because of the known evolution of the $\mathrm{M}_*$--$\mathrm{R}_\mathrm{eff}$ relationship, the lack of evolution in the MFP is not immediately obvious. Velocity dispersions tend to be higher, however, for more compact galaxies, which would suggest that younger galaxies have higher velocity dispersions and therefore higher SMBH masses.  This does not mean that black holes decrease in mass, of course, but suggests that black holes grow faster (relatively) than their host galaxies at first.  This inference is supported by observations of red nugget galaxies. We therefore investigate  how the assumption of SMBH mass galaxy scaling relation affects the inferred SMBH mass population.

The structure of this paper is as follows: In section \ref{data} we describe the data we used. Section \ref{methods} provides the details of our methods and choices of scaling relations. Section \ref{results} is where we present the results of our analysis. We discuss the implications of our results in section \ref{discussion} and then summarize our work in section \ref{summary}. Tables of our fit posterior values can be found in the appendix.  Throughout this work we adopted a WMAP9 cosmology \citep{Hinshaw_2013} where $H_0 = 69.33,$ $\Omega_\mathrm{b} = 0.0472$, and $\Omega_\mathrm{c}= 0.2408$.

\section{Data}\label{data}

\begin{figure*}
    \centering
	\includegraphics[width=\textwidth]{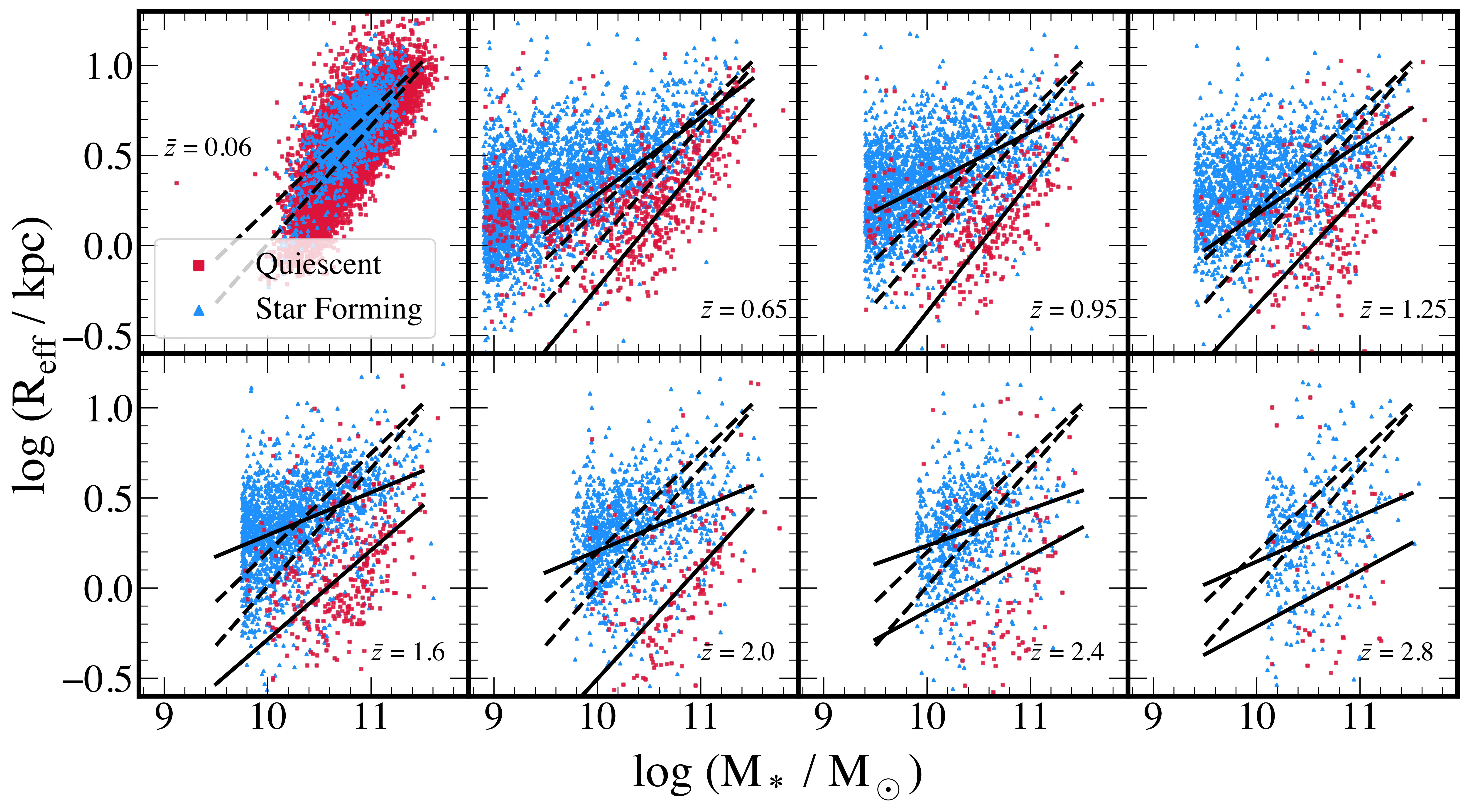}
    \caption{Mass--radius plots for quiescent (red squares) and star-forming (blue triangles)  galaxies. Each $\bar z$ represents the median redshift of the data shown in a given panel. For a fixed $\mathrm{M}_*$,  $\mathrm{R}_\mathrm{eff}$ undergoes a decrease with increasing redshift. To demonstrate this evolution, \rev{we include simple fits to the data (black lines) where dashed lines represent the lowest redshift fits for comparison in each panel. A more thorough analysis of this evolution was conducted by \citet{van_der_Wel_2014} and they report that the relation evolves as $\mathrm{R_{eff}} = 5.6 \left( \mathrm{M}_* / 5 \times 10^9 \mathrm{M}_\odot\right)^{0.8} (1 + z)^{-1.48}$ for quiescent galaxies and $\mathrm{R_{eff}} = 8.9 \left( \mathrm{M}_* / 5 \times 10^9 \mathrm{M}_\odot\right)^{0.2} (1 + z)^{-0.75}$ for star forming galaxies.} Because of this decrease, a non-evolving MFP implies different predictions from both \mmb\ and \msigma. Note, the lowest redshift bin is SDSS the rest show data from 3D HST+CANDELS.}
    \label{fig:RvsM}
\end{figure*}

The data we use in this work come from SDSS \citep{York_2000} and the 3D-HST+CANDELS survey \citep{Brammer_2012, Grogin_2011, Koekemoer_2011}. A summary of the data is presented in the mass--radius plots in Figure \ref{fig:RvsM}.

\subsection{Local Sample from SDSS}

\rev{\citet{Leja_2019sfr} did not provide mass estimates for galaxies below a redshift of 0.5 so, to supplement this, w}e compiled a sample of local galaxies with velocity dispersion measurements from the 7th data release of SDSS \citep{Abazajian_2009} at $0.05 < z < 0.07$ (top-left panel in Fig.~\ref{fig:RvsM}). All galaxies were selected from the SDSS Main Galaxy Sample \citep{Strauss_2002}, which is $\sim95\%$ complete \citep{Sohn_2017}. We cross-matched our initial sample with galaxies that had circularized half-light radii and stellar mass estimates from \citet{Simard_2011} and  \citet{Chang_2015},  respectively. Quiescent and star-forming galaxies were separated using their $u-r$ and $r-z$ colors, using the criteria in \citet{Chang_2015}.  These criteria are nearly identical to those laid out in \citet{Bezanson_2015}, and we found them to be consistent with other methods of separation based on, e.g., star formation rates. The data were selected for reliability of measurements and completeness of the sample from the SDSS DR7 database. We excluded flagged galaxies using the same criteria detailed in \citet{de_Graaff_2021}. For plotting purposes we include galaxies below $\log \left(\mathrm{M}_* / \mathrm{M}_{\odot}\right) = 10.5$ which \citet{de_Graaff_2021} removed from their sample entirely. Our sample contains 10,863 galaxies split into 1,241 star-forming and 9,622 quiescent galaxies.

\subsection{\texorpdfstring{0.5 < $z$ < 3.0 Sample from 3D-HST+CANDELS}{0.5 < z < 3 Sample from 3D-HST+CANDELS}}

For our high-redshift sample (all panels except top-left in Fig.~\ref{fig:RvsM}), we use data from the 3D-HST+CANDELS survey. For this work we infer SMBH mass from stellar mass and velocity dispersion, the latter of which can be calculated from stellar mass and half-light radius. Half-light radii used here are those determined by \citet{Skelton_2014}. Half-light radius estimates can differ when measured at one wavelength versus another so we normalized these radii to a rest frame of 5000 {\AA} following equation 2 in \citet{van_der_Wel_2014}. We circularized the radii according to $\mathrm{R}_\mathrm{eff} = \mathrm{R}_{\mathrm{hl}} q^{1/2}$ where $\mathrm{R}_{\mathrm{hl}}$ is the wavelength-corrected half-light radius and $q$ is the axis ratio reported by \citet{van_der_Wel_2014}. We also made cuts to the data according to \citet{van_der_Wel_2014} and \citet{Leja_2019sfr} based on, e.g., completion limits resulting in a sample that is $\geq 95\%$ complete \citep{Skelton_2014}.

Masses for each galaxy were determined by \cite{Leja_2019sfr} using the \texttt{Prospector} galaxy SED-fitting code \citep{Johnson_2017, Leja_2017}. \rev{In their work, \citet{van_der_Wel_2014} report that the mass-radius relationship evolves as $\mathrm{R_{eff}} = 5.6 \left( \mathrm{M}_* / 5 \times 10^9 \mathrm{M}_\odot\right)^{0.8} (1 + z)^{-1.48}$ for quiescent galaxies and $\mathrm{R_{eff}} = 8.9 \left( \mathrm{M}_* / 5 \times 10^9 \mathrm{M}_\odot\right)^{0.2} (1 + z)^{-0.75}$ for star forming galaxies. Because their analysis was performed with different mass estimates, we provide our own fits to the data to demonstrate this evolution. Those interested in the evolution of this relationship should refer to \citet{van_der_Wel_2014} for a more rigorous characterization of this relationship.} Our final sample consists of 13,232 galaxies from the UDS, GOODS-S, and COSMOS fields. For all  galaxies in this sample, \citet{Leja_2019sfr} determined star formation rates from infrared (IR) and ultraviolet (UV) luminosity. We followed their galaxy type selection criteria shown in their figure 5 resulting in a final sample of 11,107 star-forming and 2,125 quiescent galaxies.

\section{Methods}\label{methods}

Here we describe how we use the \ref{MFP} to infer velocity dispersions for all galaxies in our sample, as well as the two methods of predicting SMBH mass that are our main focus of this paper. The resulting SMBH mass predictions are converted to number density functions, the process for which is detailed at the end of this section. 

\subsection{Scaling Relations}

In this section we give the relations for the \rev{MFP}, \msigma\ and \mmb.

\subsubsection{High Redshift Velocity Dispersion}

We infer velocity dispersions for our sample using the galaxy MFP; a three-dimensional relation between galaxy stellar mass, half-light radius, and stellar velocity dispersion \citep{Hyde_2009}. This relation can be used reliably to predict any of the three properties if the other two are known. Several works in the last decade have investigated both the possibility of an evolution in the MFP and the effect galaxy type may have on the parameterization \citep{Gebhardt_2003g, Peralta_2015, Beifiori_2017}. Now, with large volumes of deep data a picture is emerging where all galaxies lie on one plane that does not evolve \citep[at least out to $z \sim 1$,][]{Bezanson_2013, Bezanson_2015, de_Graaff_2020, de_Graaff_2023}. In particular, \citet{de_Graaff_2021} recently performed a thorough analysis of the galaxy type dependence and redshift evolution and came to this same conclusion. Motivated by these results we used the MFP described described by
\begin{equation}
    \log\sigma = (\log \; \mathrm{R}_\mathrm{eff} - \beta \;\log\Sigma_{\star} - \gamma)\; / \;\alpha
	\label{MFP}
\end{equation}
and 
\begin{equation}
    \Sigma_{\star} \equiv \mathrm{M}_* \, / \,(2 \pi \mathrm{R}_\mathrm{eff}^2),
	\label{sigmastar}
\end{equation}
where $\alpha = 1.6287$ and $\beta = -0.84$ as determined by \citet{Hyde_2009} and the offset is $\gamma= 4.482$ \citep{de_Graaff_2021}.

\begin{figure}
	\includegraphics[width=\columnwidth]{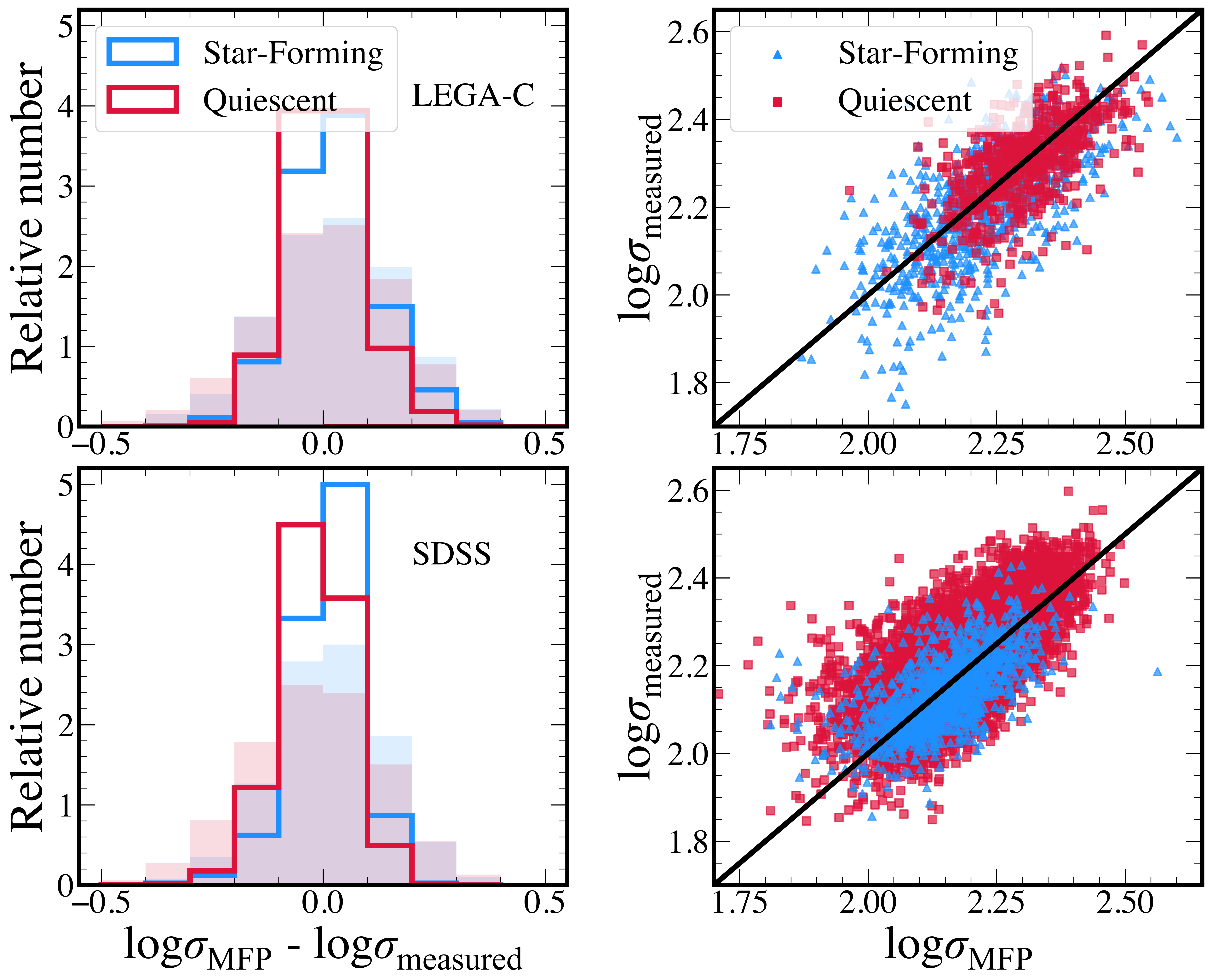}
    \caption{Residuals for velocity dispersion inferred from the MFP using both the SDSS and LEGA-C surveys. Shaded, low-alpha histograms include scatter. When including scatter the standard deviation the histograms is below 0.16 dex for all galaxies. Without intrinsic scatter included we find a standard deviation of at most 0.08 dex for quiescent galaxies and 0.1 dex for star-forming. Because our inferred velocity dispersions reproduce the measured values so well, the MFP velocity dispersions we infer can be treated functionally like measured values.}
    \label{fig:mvpsigma}
\end{figure}

If the MFP is a valid prescription, we should be able to reproduce measured velocity dispersions using the stellar mass and effective radii of each galaxy. We compare the measured velocity dispersions from galaxies in both the SDSS and LEGA-C surveys to those we predict using the MFP.  We plot the results of these comparisons in Fig.\ \ref{fig:mvpsigma} for each set of galaxies. We find that our predicted values are consistent with measurements for all galaxy types across both samples (0.1 dex or below), even with scatter introduced (0.16 dex or lower). Because our predictions are able to reproduce the measured values, we can treat the MFP velocity dispersions functionally as measured velocity dispersions. From here on we use $\sigma$ to indicate the velocity dispersion predicted from the MFP unless otherwise specified.

\subsubsection{Supermassive Black Hole Mass}

To infer SMBH mass from host galaxy properties we used the relations presented in \citet{Kormendy_Ho_2013} for the \mmb\ and \msigma\ scaling relationships given by
\begin{equation}
    \frac{\mathrm{M}_\mathrm{BH}}{10^9 M_{\odot}}=\alpha_1 \left(\frac{M_{\mathrm {bulge}}}{10^{11} \mathrm{M}_{\odot}}\right)^{\beta_1}
	\label{mm}
\end{equation}
and 
\begin{equation}
    \frac{M_\mathrm{BH}}{10^9 \mathrm{M}_{\odot}}=\alpha_2\left(\frac{\sigma}{200 \mathrm{~km} \mathrm{~s}^{-1}}\right)^{\beta_2}.
	\label{msigma}
\end{equation}
The two relations are well studied in the local universe, but there is a lack of consensus surrounding the evolution (or lack thereof) of either relation beyond nearby galaxies \citep{Croton_2006, Gaskell_2009, Mountrichas_2023, Robertson_2006, Salviander_2013, Sun_2015, Treu_2007, Woo_2006, Woo_2008, Merloni_2010, Silverman_2022, Shen_2015}. For this work we assumed the local paramtetrization [$\alpha_1$, $\beta_1$] = [0.49, 1.16] and [$\alpha_2$, $\beta_2$] = [0.309, 4.38] to be non-evolving with redshift.  We revisit this assumption in section \ref{discussion}. When using mass and radius to predict velocity dispersion, the \msigma\ relation becomes a function of both bulge mass and radius, therefore including an additional galaxy property in the mass estimation in contrast with \mmb. Because of this consideration of galactic radius, \msigma\ implicitly incorporates the evolution of the $\mathrm{M}_*$--$\mathrm{R}_\mathrm{eff}$ relationship with redshift without defining an explicit redshift evolution \citep[see also][]{van_den_Bosch_2016}.

Because SMBH mass is derived from host bulge properties, we assigned each star-forming galaxy a bulge mass fraction to be 40\% of its total stellar mass. Our choice of bulge mass fraction has an effect on the degree to which the two relationships disagree, but the our overall results do not change when using significantly higher or lower fractions. We also performed our analysis for each galaxy type separately, so results including only quiescent galaxies are not affected by this choice.

\subsection{Number Density Functions}

The stellar mass function (SMF) of galaxies is a useful tool for understanding galaxy formation and evolution. The SMF informs us of the total number of galaxies per unit volume per logarithmic mass interval as a function of stellar mass. Though stellar mass and luminosity are the most commonly discussed, this type of number density function, $\Phi(\mathrm{X})$, can be constructed for virtually any galaxy property.

There are several ways of estimating $\Phi(\mathrm{X})$, but the most straightforward is Schmidt's $1 / \mathrm{V}_\mathrm{max}$ method \citep{Schmidt_1968, Avni_1980}. We calculate the density functions as
\begin{equation}
    V_{\max , i}=\frac{\Omega}{3}\left(r\left(z_{\max , i}\right)^3-r\left(z_{\min , i}\right)^3\right)
	\label{vmax}
\end{equation}
and
\begin{equation}
    \Phi(\mathrm{X}) =\sum_i \frac{1}{\mathrm{V}_{\max , i}  \Delta \mathrm{X}},
	\label{oneovervmax}
\end{equation}
where $\mathrm{X}$ represents the property in question, e.g., stellar mass, velocity dispersion, or SMBH mass and $\mathrm{V}_{\max , i}$ is the co-moving volume between redshifts $z_{\min , i}$ and $z_{\max , i}$. The solid angle subtended by the survey is represented by $\Omega$, and $\Delta \mathrm{X}$ is the width of the bins. This method is functionally similar to a histogram making it computationally efficient and it is robust against bias as long as no clustering is present \citep{Marchesini_2007}. Given the high completeness of the data sets we use, this is sufficient for our purposes.
Because $\Phi(\mathrm{X})$ is a function of redshift, it is common to split the data into narrow redshift bins and fit each independently. We used the survey areas listed in \citet{Skelton_2014} to calculate our co-moving volume for each redshift bin.

The number of galaxies within a given volume is expected to undergo an overall decline with increasing redshift and with increasing extremity of the property in question (e.g., very high mass or luminosity). Distributions of $\Phi(\mathrm{X})$ of this sort are well described by Schechter functions. The logarithmic form of a ``single Schechter'', which we used for all our fitting, is described by
\begin{equation}
   \Phi(\mathrm{Y})=\ln (10) \phi_* 10^{\left(\mathrm{Y}-\mathrm{Y}_\mathrm{c}\right)(\alpha_s+1)} \exp \left(-10^{\mathrm{Y}-\mathrm{Y}_\mathrm{c}}\right),
	\label{singleschechter}
\end{equation}
where Y is the base 10 logarithm of the property in question, i.e. Y = log$_{10}$(X),  $\mathrm{Y}_\mathrm{c}$ is the (log) characteristic value of said property, $\alpha_s$ is the slope of the lower power-law, and $\phi_*$ is density normalization. Especially \rev{in the local universe}, a ``double Schechter'' is sometimes used which is simply the sum of two single Schechter functions.

\begin{figure}
	\includegraphics[width=\columnwidth]{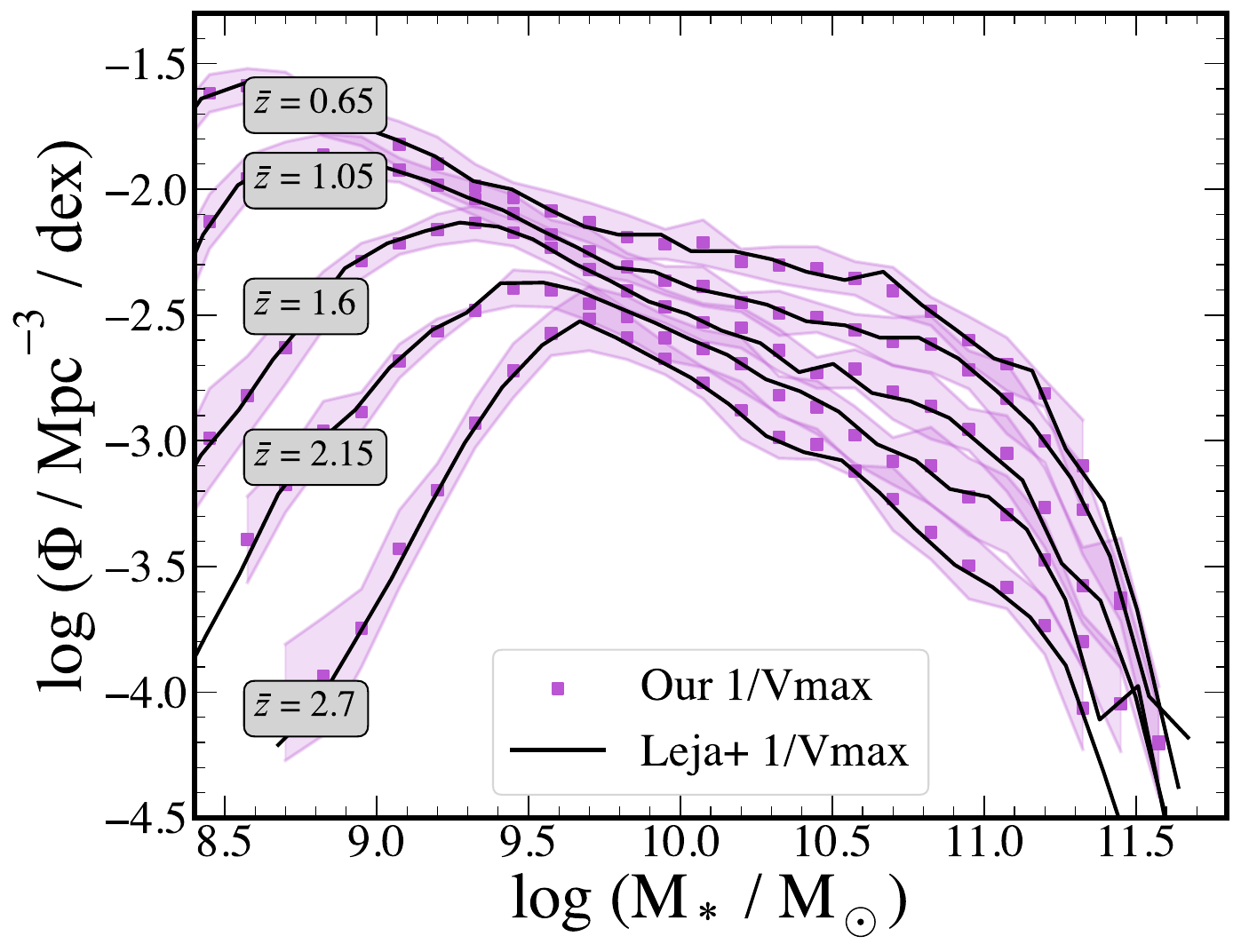}
    \caption{Our SMFs compared with those in \citet{Leja_2019}. The solid black lines are their $1 / \mathrm{V}_\mathrm{max}$ estimates and the purple squares are our own. Purple shaded regions represent the error in our $1 / \mathrm{V}_\mathrm{max}$ estimates, it is clear at every point that our data are in good agreement with their published values.}
    \label{fig:minevsleja}
\end{figure}

After obtaining values for our stellar mass functions, we compared our estimates to those obtained in \citet[see their figure 5]{Leja_2019}. We compiled the data into one figure and over plotted our SMF estimates and found that we were in good agreement (Fig. \ref{fig:minevsleja}).

We repeated the same process to produce number density functions for velocity dispersion and SMBH mass predicted from both \mmb\ and \msigma. Our parameterization for the Schechter fits was found using PyMC \citep{Salvatier_2016}, a modeling software that uses Markov chain Monte Carlo sampling. The priors we used are listed in Table \ref{priors}. We used four chains with 15,000 total steps, the first 5,000 of which were tuning steps. In all cases, the data were not fitted for values below the completion limits.  We determined our completion limits for stellar mass from \citet{Leja_2019} and converted these into SMBH mass completion limits using the \mmb\ relation. Velocity dispersion completion limits are informed by the aforementioned limits on stellar mass and the completion limits for effective radius used by \citet{van_der_Wel_2014}. A more complete breakdown can be found in Table \ref{comlim}.

\begin{table}
\begin{center}
\begin{tabular}{|cccc|}
\hline
Median Redshift & log M$_*$ [M$_\odot$] & log $\sigma$ [km s$^{-1}$] & log M$_\mathrm{BH}$ [M$_\odot$]\\
\hline
0.65  & 9.1 & 2.2 & 6.25 \\
0.95  & 9.5 & 2.2 & 6.5 \\
1.25  & 9.7 & 2.2 & 6.5 \\
1.60  & 9.9 & 2.4 & 7.0 \\
2.00  & 10.0 & 2.4 & 7.0 \\
2.40  & 10.2 & 2.4 & 7.5 \\
2.80  & 10.2 & 2.3 & 7.5 \\
\end{tabular}
\caption{\label{comlim} Completion limits for stellar mass, velocity dispersion, and SMBH mass. Values greater than those listed in this table are part of the complete sample and are considered reliable.}
\end{center}
\end{table}

Error estimates were obtained by performing 100 fits to the data where we introduced random scatter into the data based on the errors of the values involved in the fits and the known intrinsic scatter of the relations used for our inferred quantities. Cosmic variance estimates were obtained following the methods outlined in \citet{Moster_2011}. Because accurate determinations of cosmic variance for velocity dispersion and SMBH mass would require a large volume of in-depth measurements for each of these values, an exact estimate does not exist. For these values we approximated the cosmic variance based on the values we calculated for stellar mass.

\begin{table}
\begin{center}
\begin{tabular}{|cc|}
\hline
Stellar Mass Parameters & Prior Bounds \\
\hline
$\log \phi_{*, 1}$             & $-6$, $-2$ \\
$\log \phi_{*, 2}$             & $-6$, $-2$ \\
$\alpha_{s, 1}$                & $-1$, $1$ \\
$\alpha_{s, 2}$                & $-2$, $-1$ \\
log $\mathrm{M}_\mathrm{c}$    & $10$, $12$ \\
$\log \sigma_\mathrm{scatter}$ & $-2$, $-0.5$ \\
\hline
Velocity Dispersion Parameters & Prior Bounds \\
\hline
$\log \phi_*$                  & $-8$, $-3$ \\
$\alpha_s$                     & $3$, $8$ \\
log $\sigma_\mathrm{c}$        & $0$, $2.5$ \\
$\log \sigma_\mathrm{scatter}$ & $-2$, $-0.5$ \\
\hline
Black Hole Mass Parameters & Prior Bounds \\
\hline
$\log \phi_*$                  & $-6$, $-2$ \\
$\alpha_s$                     & $-4$, $4$ \\
log $\mathrm{M}_\mathrm{c}$    & $5$, $12$ \\
$\log \sigma_\mathrm{scatter}$ & $-2$, $-0.5$ \\
\hline
\end{tabular}
\caption{\label{priors}Prior ranges for Schechter function fits to the data. We used uniform distributions between the values listed.}
\end{center}
\end{table}

\section{Results}\label{results}

In Figures \ref{fig:SMF8Panel}, \ref{fig:SMFOnePanel}, \ref{fig:VDF8Panel}, and \ref{fig:VDFOnePanel} we present the number density functions of galaxy stellar mass, MFP velocity dispersion, and inferred SMBH mass from both the \mmb\ and \msigma\ scaling relations.

\subsection{Stellar Mass and Velocity Dispersion Functions}

\begin{figure*}
\centering
	\includegraphics[width=\textwidth]{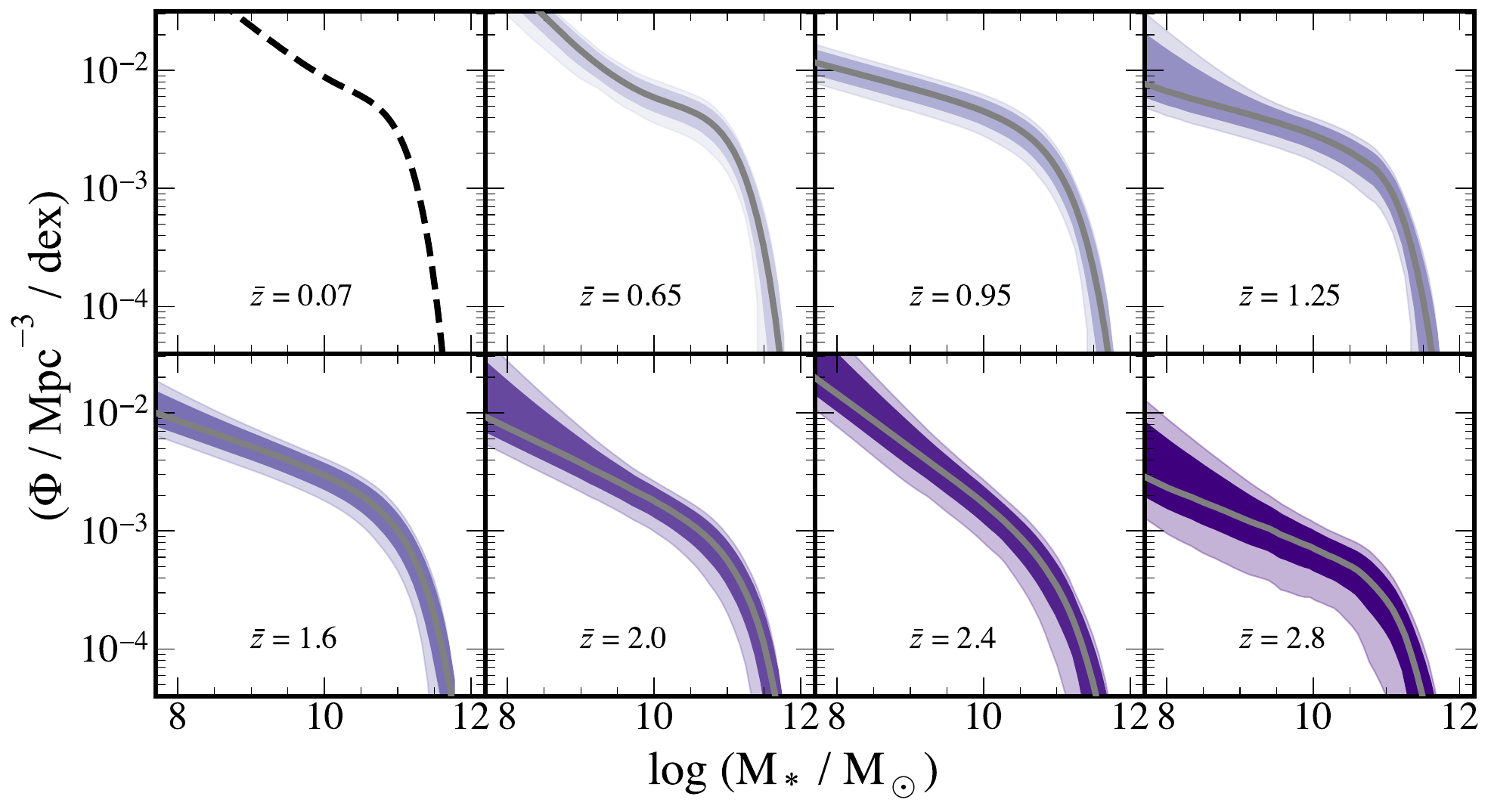}
    \caption{All Schechter fits to the stellar mass functions across all redshifts. We fit a double Schechter for all the functions shown here which can be seen in the double sloped nature of the lower mass end of some distributions. Each $\bar z$ represents the median redshift of the data shown in a given panel. The solid lines through each curve represents the median value to these fits and the shaded regions are our 68\% an 95\% confidence limits plus cosmic variance for the darker and lighter colors respectively. We do not find any significant evolution between panels other than the general decrease in number density at all masses as redshift increases. The dashed line in the first panel is the result  for galaxies at low redshift calculated from the method provided in \citet{Leja_2019}.}
    \label{fig:SMF8Panel}
\end{figure*}

\begin{figure}
	\includegraphics[width=\columnwidth]{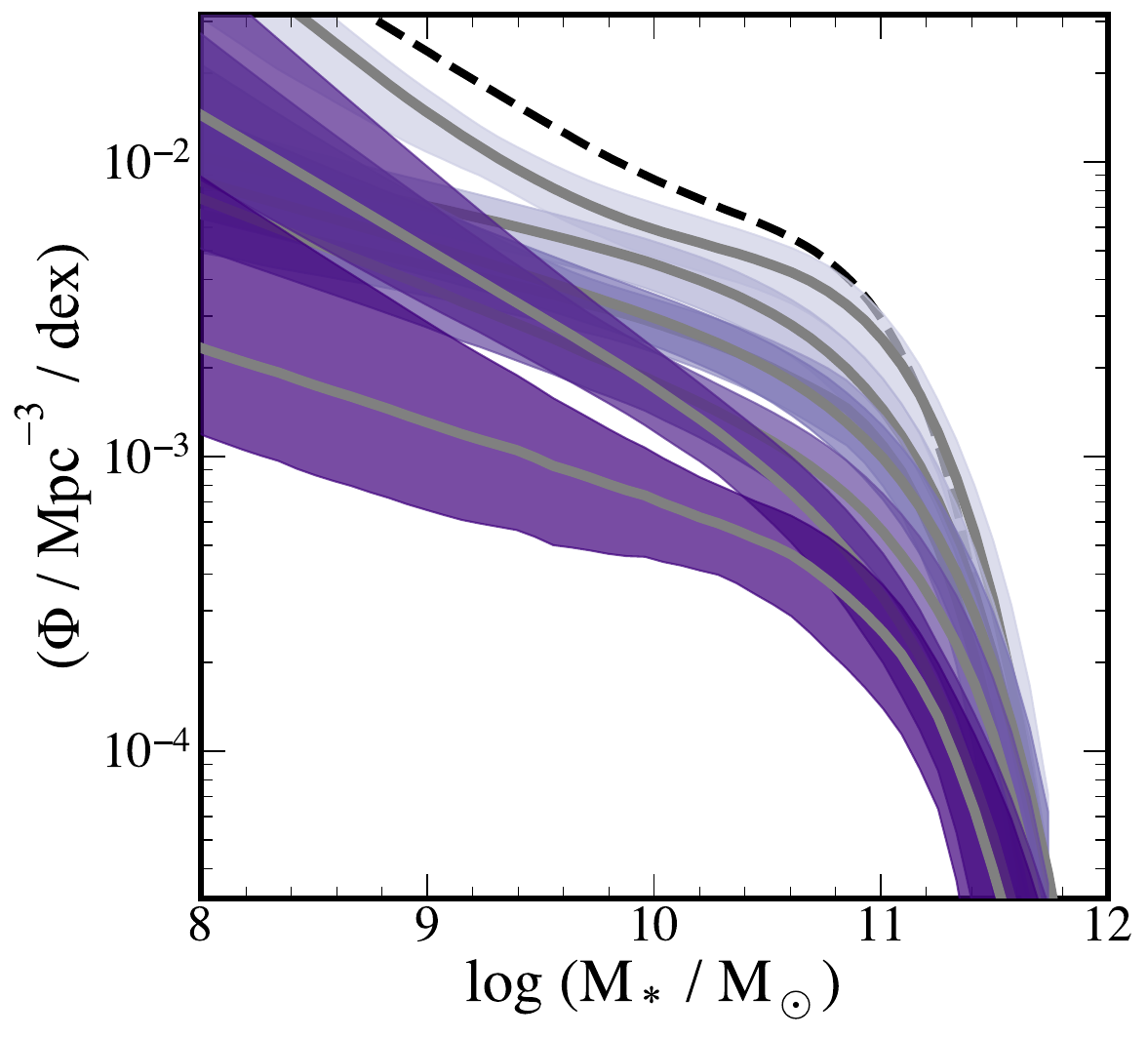}
    \caption{Same as in Figure \ref{fig:SMF8Panel} but all redshifts are shown. Other than the overall decrease, we do not notice any substantial trends across time. The colors at each redshift are the same here where generally as redshift increases the curves appear lower on the plot.}
    \label{fig:SMFOnePanel}
\end{figure}

\begin{figure*}
\centering
	\includegraphics[width=\textwidth]{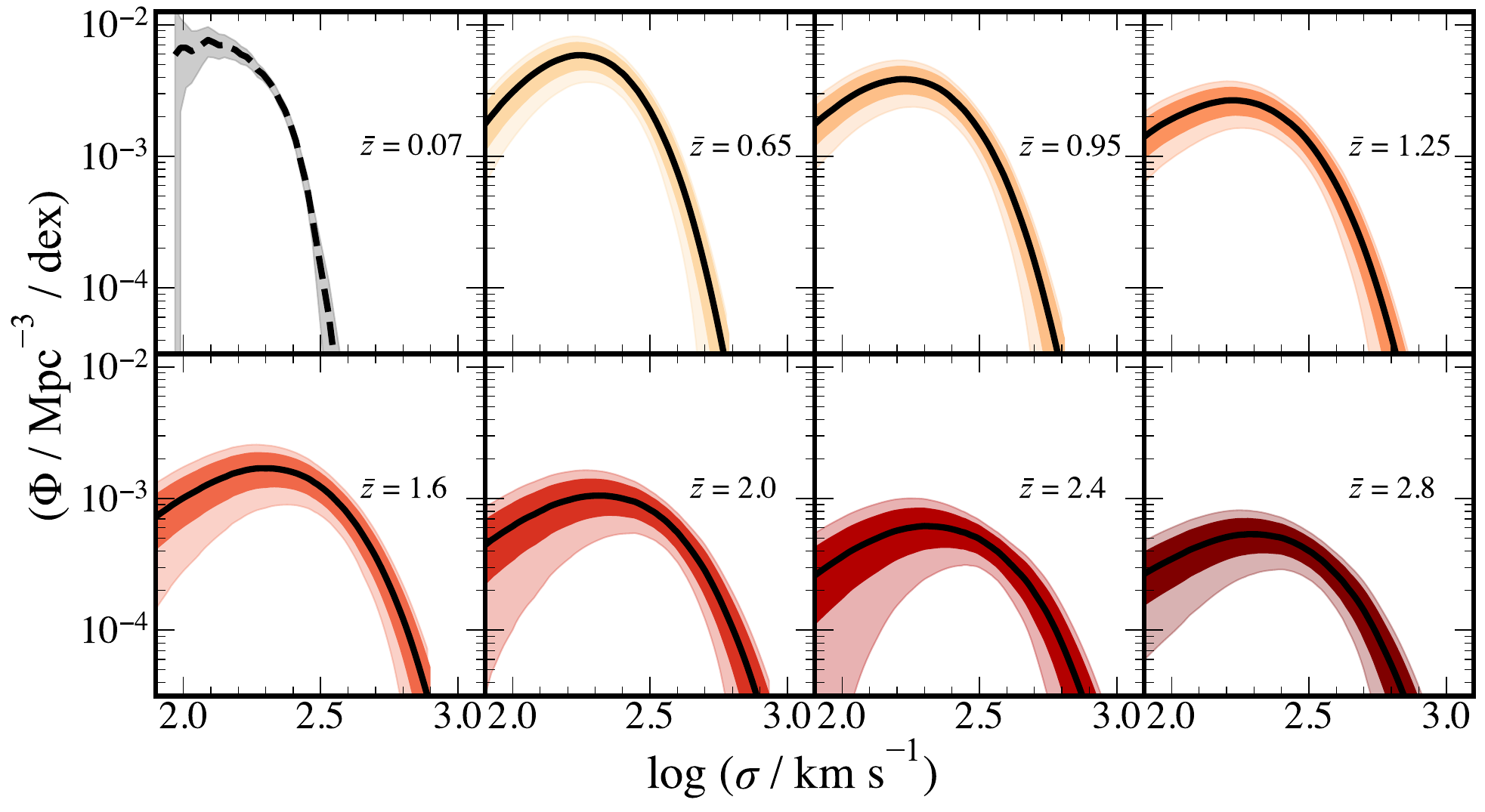}
    \caption{All Schechter fits to the velocity dispersion functions across all redshifts. The data here are characterized by single Schechter functions. Each $\bar z$ represents the median redshift of the data shown in a given panel. The solid lines through each curve represents the median value to these fits and the shaded regions are our 68\% an 95\% confidence limits plus cosmic variance for the darker and lighter colors respectively. We note the decrease in number density with increasing redshift across the entire range of velocity dispersions. We also see a trend of increasing characteristic velocity dispersion with increasing redshift. The dashed line in the first panel is the result from \citet{Sohn_2017} for quiescent galaxies in SDSS for 0.03 < $z$ < 0.1.}
    \label{fig:VDF8Panel}
\end{figure*}

\begin{figure}
	\includegraphics[width=\columnwidth]{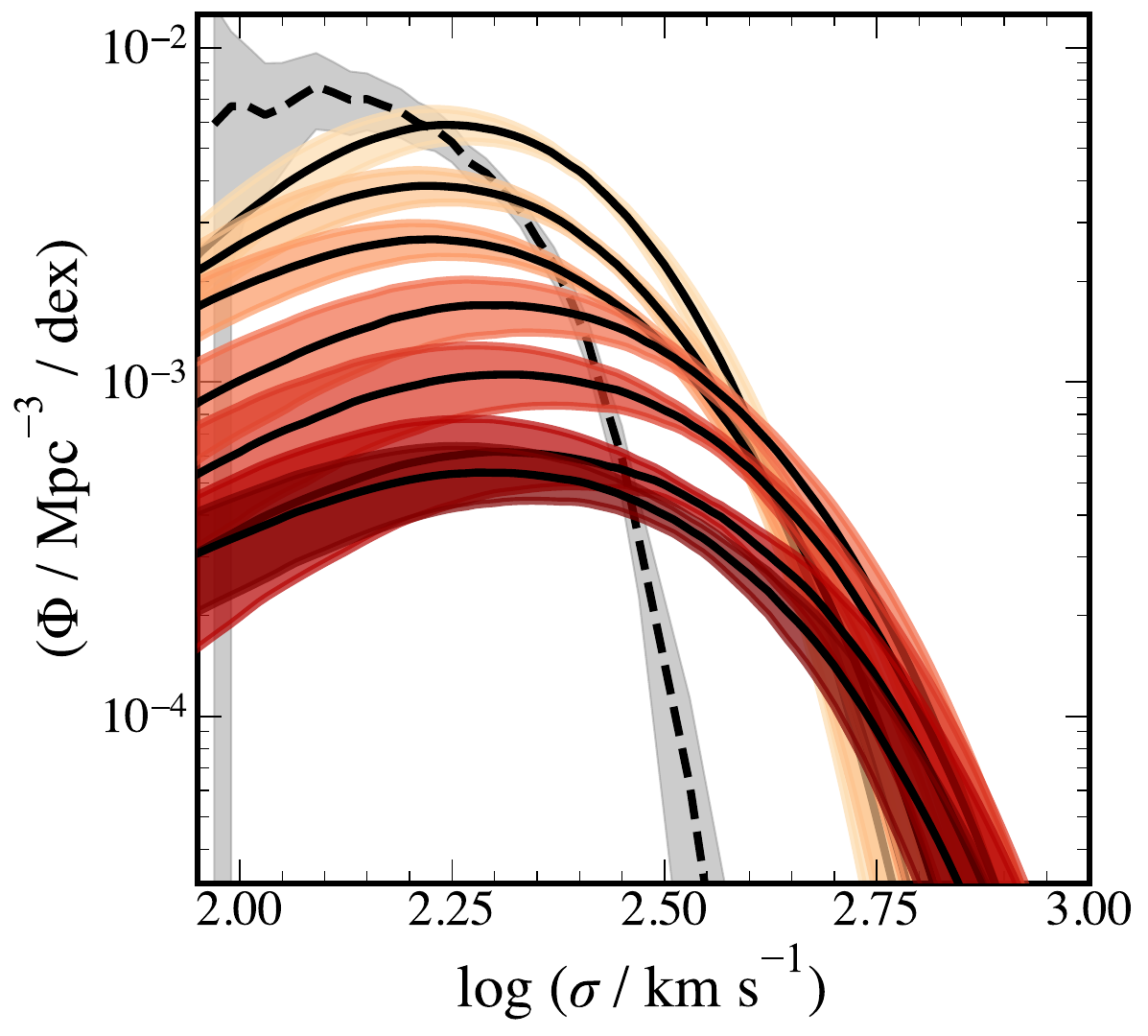}
    \caption{Same as in Figure \ref{fig:VDF8Panel} but all redshifts are shown. The colors at each redshift are the same here where generally redshift is increasing as the curves move down the plot. The large gap between the results of \citet{Sohn_2017} and our functions has several likely origins including the large jump in cosmic time ($\sim$ 5.2 Gyr) between their results and our lowest redshift, and the variance in the scaling relations we used. This is discussed further in the text.}
    \label{fig:VDFOnePanel}
\end{figure}

Our stellar mass and velocity dispersion function fits to all galaxies are shown in figures  \ref{fig:SMF8Panel}--\ref{fig:VDFOnePanel}, . The stellar mass functions (Figs.\ \ref{fig:SMF8Panel} and \ref{fig:SMFOnePanel}) are described here by a double Schechter function at all redshifts. At the highest redshifts the data are well described by a single Schechter which is consistent with others' results \citep[e.g.,][]{McLeod_2021}, but we chose to fit these with a double Schechter to maintain consistency within our results across all redshifts. There is a general decline in the total number density between the lowest and highest redshifts, the number of galaxies with $\log{\mathrm{M}_*} > 11.5\  \mathrm{M}_\odot$ is 8.3 times higher at $\bar z = 0.65$ than at $\bar z = 2.8$. The distribution, $\Phi(\mathrm{M}_*)$ drops off steeply for masses greater than $\sim 11\ \mathrm{M}_\odot$ but the slope for lower masses is much flatter with no clear trends across time. 

The velocity dispersion functions (Figs.\ \ref{fig:VDF8Panel} and \ref{fig:VDFOnePanel}) are parameterized by a single Schechter function across all redshifts. We see an overall decrease in number density of galaxies as redshift increases. There appears to be a mild change in the slope of the distribution that is steepest at $\bar z = 0.65$ and is at its  shallowest for $1.6 < \bar z < 2.0$. This flattening of the curve leads to an apparent broadening of the whole distribution, though we cannot be sure if the flattening of the values to the left of the completion limits are reliable. Perhaps the most notable results of these fits are the evolution of the characteristic velocity dispersion which increases from 1.6 to 1.9 over the entire redshift range. An increase of the characteristic velocity dispersion suggests that galaxy velocity dispersion is increasing with increasing redshift.

The large difference between the results of \citet{Sohn_2017} and our functions (Fig. \ref{fig:VDFOnePanel}) has several possible explanations. First, their results consider only quiescent galaxies while ours are for combined galaxy type. Number density functions of separate galaxy types often have different shapes to the combined functions as we find in this paper and what was found by, e.g., \citet[see also \citealt{Bezanson_2011}]{Taylor_2022}. There is also a large gap in cosmic time between their $\bar z = 0.07$ results and our lowest redshift sample which is $\bar z$ = 0.65 that corresponds to a approximately 5.2 Gyr. Because we see lower characteristic velocity dispersions with lower redshift, it is possible that the relation evolves in this time. Additionally, \citet{Bezanson_2011} found an increase in the number of galaxies with high velocity dispersions for $z > 0.6$ which could indicate an evolution in the intrinsic scatter of the relation they used to infer velocity dispersion. Though they used dynamical mass to infer virial velocity dispersions, which is different to what we do here, a similar scatter evolution could be affecting this difference since we include the measured intrinsic scatter from \citet{de_Graaff_2020} which was measured for $z \sim 0.8$.

\subsection{Supermassive Black Hole Mass Functions}

\begin{figure*}
    \centering
	\includegraphics[width=\textwidth]{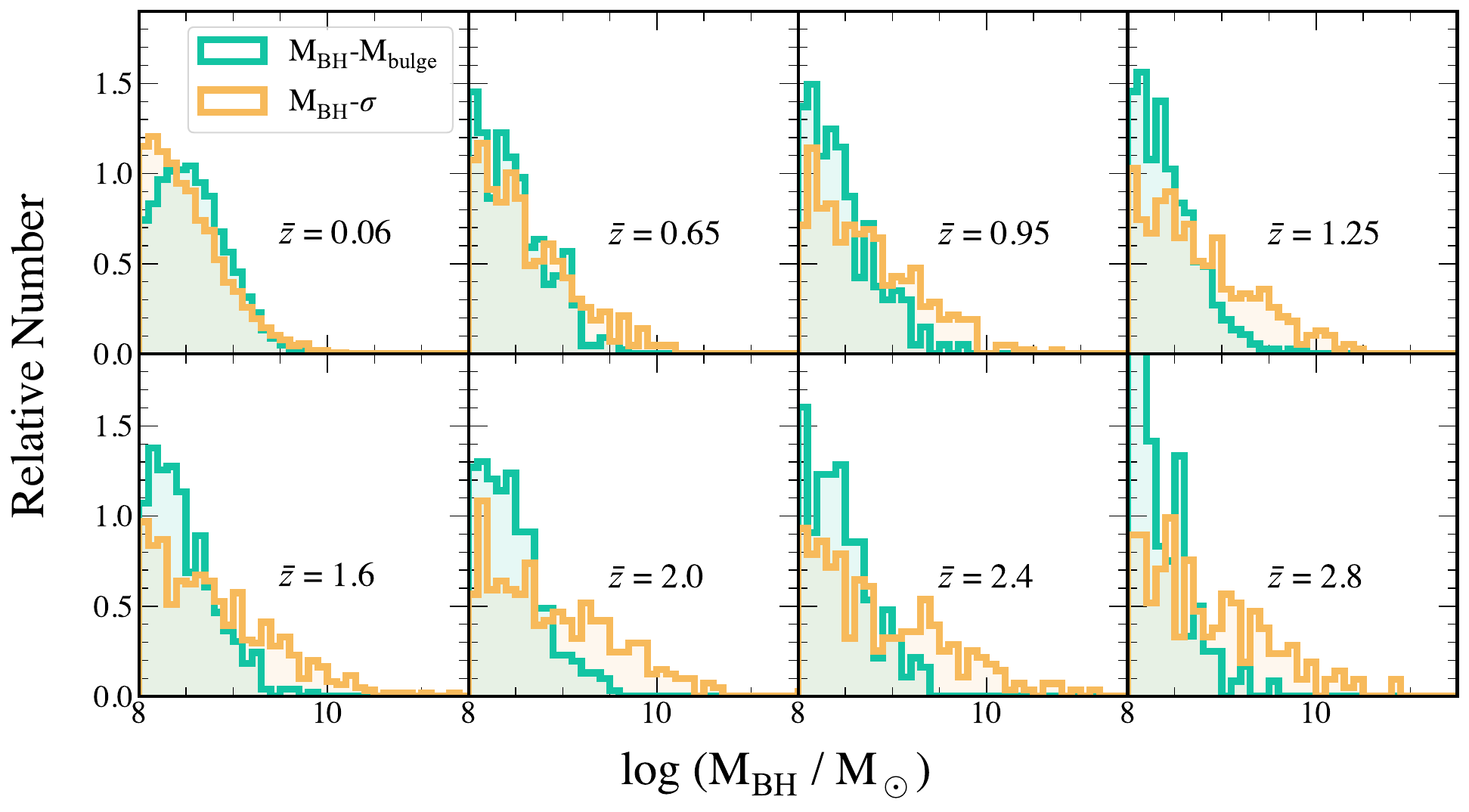}
    \caption{The relative numbers of SMBH masses as predicted by the \mmb\ (green) and the \msigma\ (orange) relationships for all galaxy types. Each $\bar z$ represents the median redshift of the data shown in a given panel. While the two distributions are very similar in shape, especially at high masses, for nearby galaxies, the \msigma\ tail dominates for high redshift galaxies with predicted SMBH masses above $\sim 10^9\ \mathrm{M}_\odot$.}
    \label{fig:histograms}
\end{figure*}

\begin{figure*}
    \centering
	\includegraphics[width=\textwidth]{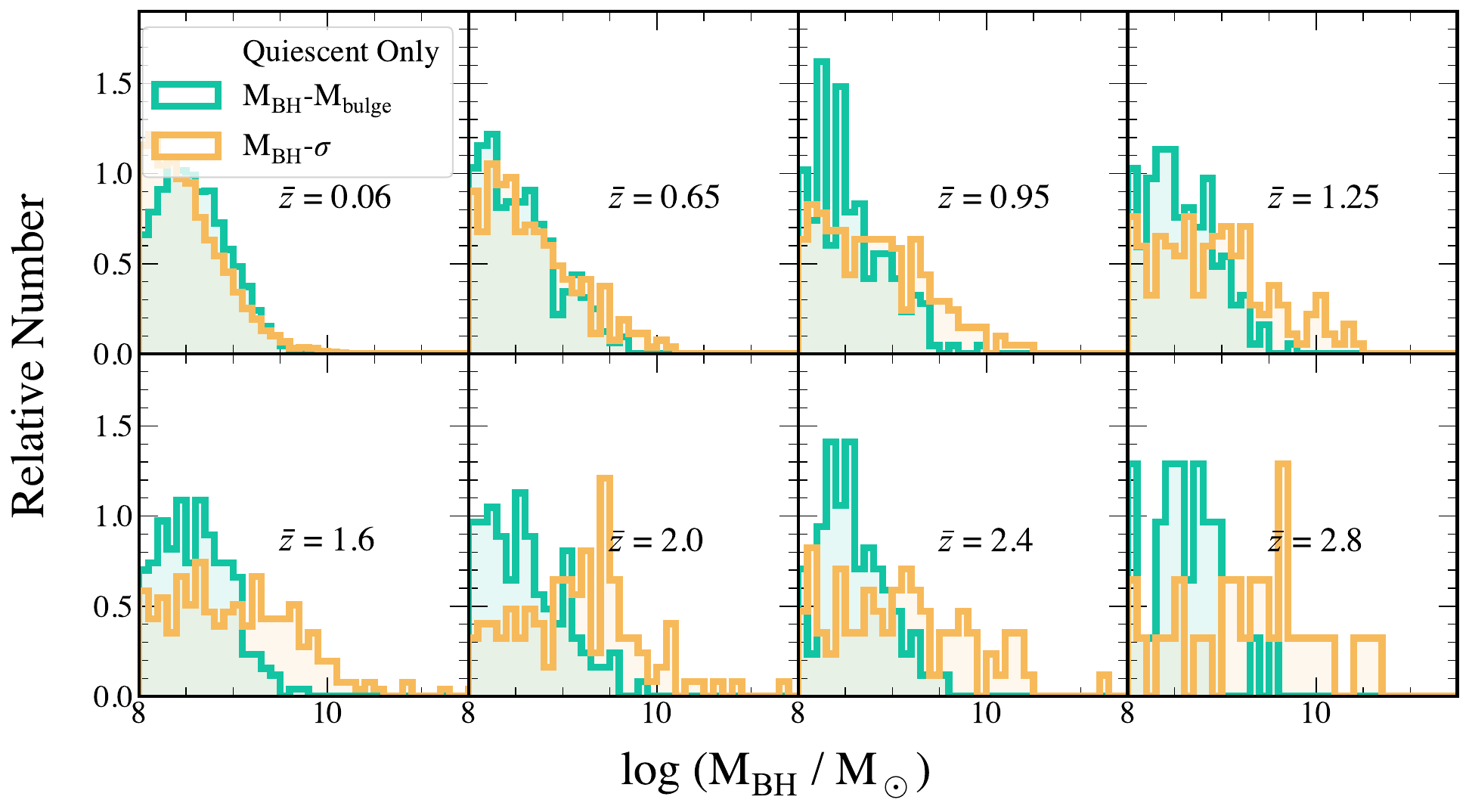}
    \caption{Same as Figure \ref{fig:histograms}, but only quiescent galaxies are shown. The discrepancy in predicted SMBH masses is more pronounced when we consider the quiescent population independently. The fact that we predict significantly different distributions of SMBH masses when using velocity dispersion versus\ stellar mass in this quiescent-only sample reinforces that our results are not biased by our choice of bulge mass fraction for star-forming galaxies.}
    \label{fig:histogramsEARLY}
\end{figure*}

We show histograms of resulting distributions of SMBH masses in Figure \ref{fig:histograms}. As we look back to earlier times the shape of the histogram of SMBH masses inferred from velocity dispersions flattens out leading to a lower peak, but a much thicker and longer tail than for SMBH masses inferred from stellar mass. These same data are shown in Figure \ref{fig:histogramsEARLY} showing only our quiescent galaxy population. We see the same trends here despite having far fewer galaxies; the high mass tail of the distribution is larger for masses predicted from velocity dispersion than from stellar mass. It is from these same data that we constructed the mass functions for each relationship for star-forming, quiescent, and combined galaxy types. 

\begin{figure}
	\includegraphics[width=\columnwidth]{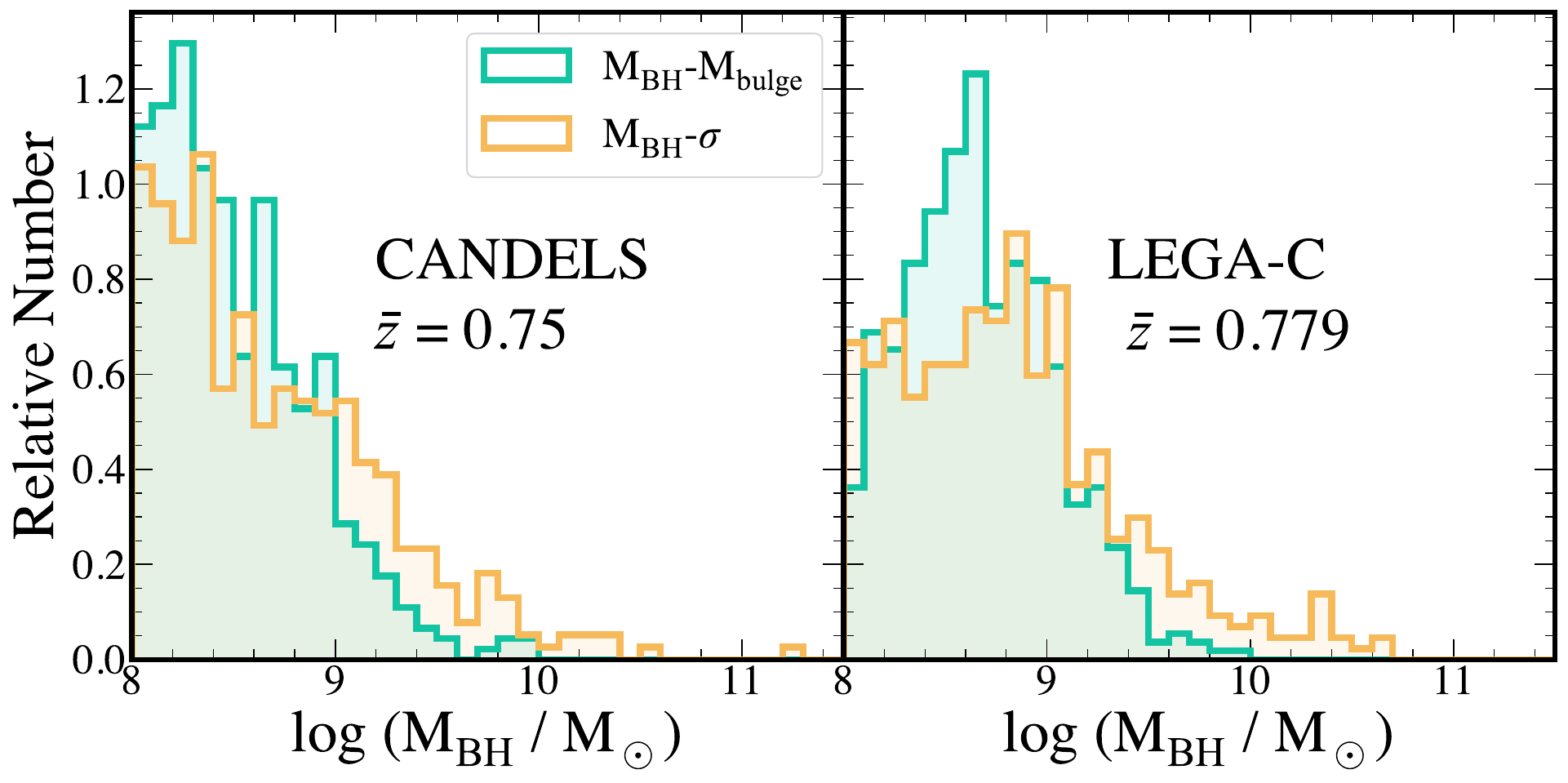}
    \caption{Comparison between SMBH mass predictions for quiescent galaxies in 3D HST+CANDELS and LEGA-C. The median redshift, $z$ is shown for each survey in the plot. We see similarities between the two distributions reinforcing that the higher numbers of high mass SMBHs inferred from velocity dispersion is not an artifact of sample choice.}
    \label{fig:legvscand}
\end{figure}

If our results are to be trusted, they should be independent of survey choice. We can compare CANDELS to the LEGA-C survey for quiescent galaxies between $0.5 \lesssim z \lesssim 1$.  In this redshift range, the two surveys have comparable coverage, and even though our results are robust to choice of bulge fraction, we see these same results even when restricting to quiescent galaxies only. When repeating our analysis on LEGA-C (Fig.\ \ref{fig:legvscand}), we get SMBH mass distributions that have all of the same properties we have highlighted. Namely, \msigma\ predicts a larger number of SMBHs with masses greater than $\sim 10^9\ \mathrm{M}_\odot$ and also extends to higher masses than \mmb. The fact that we find similar trends between both data sets with quiescent galaxies suggests that our results are both reproducible and unbiased by survey choice or bulge stellar mass fraction.

\begin{figure*}
	\includegraphics[width=\textwidth]{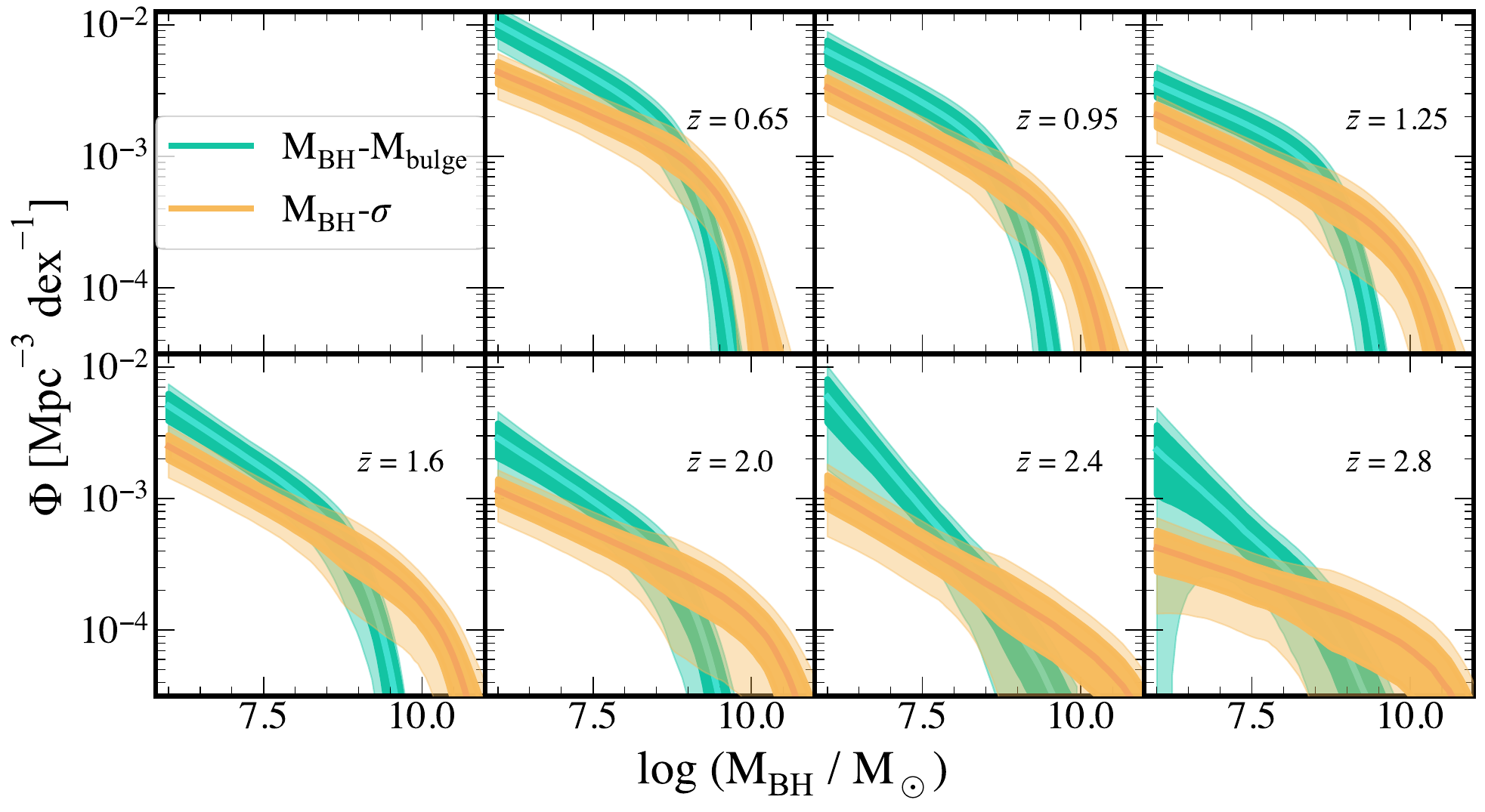}
    \caption{Single Schechter fits to SMBH mass functions predicted by the \mmb\ (green) and the \msigma\ (orange) for all galaxy types. Each $\bar z$ represents the median redshift of the data shown in a given panel. We see that the two relations differ at both the low and high mass regions of each distribution. It is clear that the number density of the highest mass SMBHs is much greater when using velocity dispersion to infer their masses as opposed to stellar mass. The area of each plot around the lines represent the 68\% (darker) and 95\% (lighter) confidence intervals plus cosmic variance.}
    \label{fig:bhmf}
\end{figure*}

\begin{figure*}
    \centering
	\includegraphics[width=\textwidth]{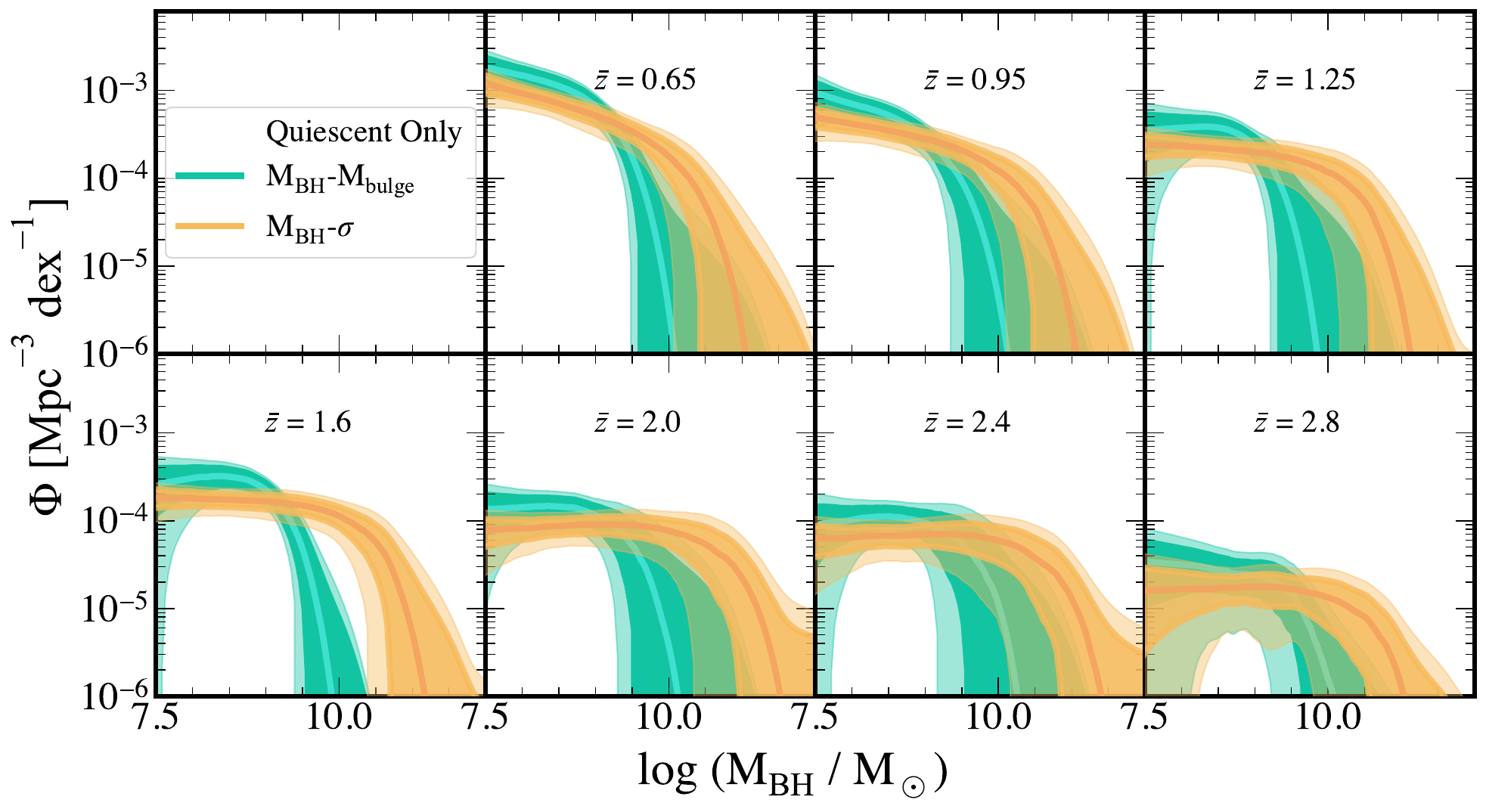}
    \caption{Same as Figure \ref{fig:bhmf}, but only quiescent galaxies are shown. We see the same general differences between the masses predicted from each of stellar mass and velocity dispersion with the latter producing more SMBHs at the higher mass end. The solid lines through each curve represents the median value to these fits and the shaded regions are our 68\% an 95\% confidence limits plus cosmic variance for the darker and lighter colors respectively. There is significantly more overlap within our errors here though the median fits remain separated.}
    \label{fig:mmvsmsfuncearly}
\end{figure*}

\begin{figure*}
    \centering
	\includegraphics[width=\textwidth]{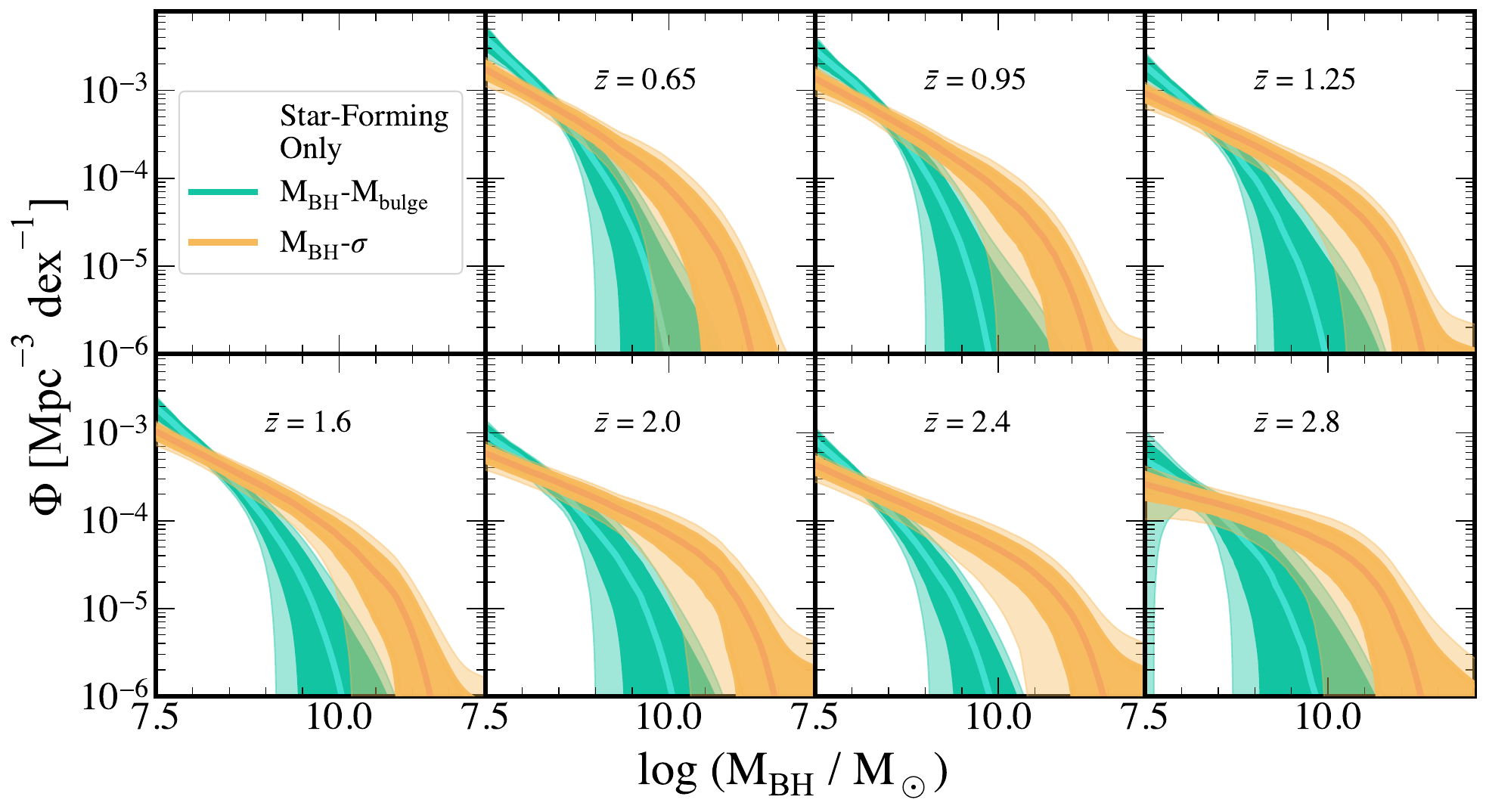}
    \caption{Same as Figure \ref{fig:bhmf}, but only star-forming galaxies are shown. As before, the median fits are represented by solid lines and the shaded regions show our 68\% an 95\% confidence limits plus cosmic variance for the darker and lighter colors respectively. The distributions here differ more significantly than either of the quiescent-only and all galaxy type plots. This increased disagreement may be impacted by our choice of bulge mass fraction when isolating bulge mass for predictions from \mmb.}
    \label{fig:mmvsmsfunclate}
\end{figure*}

The resulting SMBH mass functions for both galaxy types as well as quiescent and star-forming galaxies are shown in Figures \ref{fig:bhmf}, \ref{fig:mmvsmsfuncearly}, \ref{fig:mmvsmsfunclate} respectively. Here median fits and errors are presented in the same way as the stellar mass and velocity dispersion fit. We find that, independent of galaxy type, there are significant differences between the predicted SMBH masses from \msigma\ and \mmb\  especially \rev{for redshifts above 1}. For all redshift bins higher than $z \sim 1$, \msigma\ predicts a notably higher number density of large (M$_\mathrm{BH} > 10^9\ \mathrm{M}_{\odot}$) SMBHs. While both relationships undergo a decrease in total number density with increasing redshift, the overall predictions between high and low masses evolve. The number density of the highest mass black holes derived from stellar mass does not change significantly. The slope of the distribution around $\mathrm{M}_\mathrm{BH} \sim 10^8\ \mathrm{M}_\odot$ and higher remains consistent across all snapshots until a slight flattening in the two highest redshift bins. The characteristic logarithmic SMBH mass is also highest at these two times while it does not follow a noticeable trend in either direction for redshifts below $\bar z = 2.5$. The characteristic logarithmic SMBH mass for those derived from velocity dispersion undergoes an increase from 9.8 to 10.8 over the range of redshifts considered here. This change is related to the similar increase we see in characteristic velocity dispersion. The highest SMBH masses in this distribution tend towards higher values with increasing redshift which leads to a growing division further back in time.

Especially at $z \sim 3$, the distributions of SMBH masses inferred by either galaxy stellar mass or velocity dispersion do not agree. This tension is apparent when considering galaxy types both separately and together and is present across at least two different high-redshift samples (Fig. \ref{fig:legvscand}). The bulk of the distributions overlap (Fig. \ref{fig:histograms}) and so these relationships are suggesting similar populations of SMBHs for the majority of galaxies. The amplitude of the GWB is most impacted by the largest SMBHs, where the distributions differ most significantly, so an accurate picture of the high-mass population is necessary. Further study and high redshift tests of the MFP are needed.

\section{Discussion}\label{discussion}

We derive the distribution of SMBH mass for $0 < z < 3$. The masses we used were inferred from either the host bulge stellar mass or velocity dispersion, the latter being inferred from host stellar mass and radius using the MFP. When comparing these mass distributions we find that using MFP velocity dispersion implies a greater number density of SMBHs at the high mass end, particularly for $\mathrm{M}_\mathrm{BH} > 10^9\ \mathrm{M}_\odot$.

Throughout the course of this work we checked our methods against others (Figs.\ \ref{fig:RvsM}, \ref{fig:mvpsigma}, \ref{fig:legvscand}) and we were able to consistently reproduce their results and/or measured values. We additionally demonstrated that our results are not limited or biased by our choice in sample. Because higher numbers of high-mass SMBHs are predicted by \msigma\ even when only considering quiescent galaxies, we can also be confident that our choice in bulge fraction is not the reason for this this difference. Additionally, these results are not sensitive to which version of the SMBH mass scaling relationship is used. When comparing to other forms of these relations such as those determined by \citet{Gultekin_2009} or \citet{Mcconnell_Ma_2013} we found no significant differences in respective SMBH mass distributions. Finally, assuming larger values for the intrinsic scatter in the MFP and SMBH mass relations does not impact our predicted values without assuming non-physically large scatter.

Given the known observed evolution of galaxy properties, it is not possible for the $z = 0$ \mmb\ and \msigma\ relations to be both correct and non-evolving at high redshift. There have been observational studies to investigate the evolution of black hole scaling relations with sometimes contradictory results \citep{Croton_2006, Gaskell_2009, Mountrichas_2023, Robertson_2006, Salviander_2013, Sun_2015, Treu_2007, Woo_2006, Woo_2008, Merloni_2010, Silverman_2022, Shen_2015}. A recent study by \cite{Zhang_2023} uses results from HETDEX, and takes into account a number of potential observational biases including the potential selection bias discussed in \citet{Lauer_2007_bias}; they find a $0.52 \pm 0.14$ dex offset between the local \mmb\ relation and the relation at $z \sim 2$. This alone, however, does not entirely bridge the gap we find at $z \sim 2$ though their results primarily consider SMBHs with masses lower than $10^9 \mathrm{M}_\odot$ so the applicability of their results is limited when comparing to the population of large SMBHs we discuss here. Very little analysis has been performed for \msigma\ in this manner though \citet{Shen_2015} found no evolution in \msigma\ using observational data out to $z \sim 1$. Without a high redshift survey of velocity dispersions for galaxies with known SMBH mass, we have extremely limited insight into how this relation may or may not evolve.

If the observed lack of evolution in the MFP out to redshift 1 is a robust result, we would expect that any evolution in the MFP velocity dispersions out to this same redshift would reflect a physical reality. Because we see an increasing difference between the distribution of SMBH masses predicted from bulge mass and velocity dispersion even below $z = 1$, it is likely that this change is because one (or both) of these scaling relationships evolve with redshift.

We find an inescapable tension between predictions made with \mmb\ versus \msigma\ that cannot be otherwise explained given our modest assumptions. This difference in number density of high mass SMBHs has several implications for predictions such as for the sizes of galactic core. Galaxies with more massive central SMBHs have larger cores \citep{Kormendy_Ho_2013, Merritt_2006} and so using \msigma\ may predict a population of galaxies with larger cores than when using \mmb.

\rev{Our results indicate that analysis similar to \citet{Simon_2023} would point to a larger GWB amplitude when using \msigma. For masses above $10^9\ \mathrm{M}_\odot$ we can do an approximate calculation for the GWB amplitude suggested by these number densities. Following the relation between number density and GWB amplitude given in equation  \eqref{hc} we see that the amplitude has a dependence on number density such that $h_c \propto N_0^{1/2}$. Using this we can get that the ratio in amplitudes predicted by \msigma\ versus \mmb\ is proportional to the square root of the number densities of SMBHs predicted from each relation, i.e., 
\begin{equation}
   \frac{h_c(\sigma)}{h_c(\mathrm{M_{bulge}})} = \sqrt{\frac{N_0(\sigma)}{N_0(\mathrm{M}_\mathrm{bulge})}}.
	\label{hrat}
\end{equation}
Using our reported number densities (Table \ref{amplitude}) we find that using \msigma\ implies a higher amplitude by a factor of 2.1 on average across $0.5 < z < 3.0$.}

\begin{table}
\begin{center}
\begin{tabular}{|cccc|}
\hline
Median Redshift  & $N_0(\sigma)$ & $N_0(\mathrm{M}_\mathrm{bulge})$ & $h_c(\sigma) / h_c(\mathrm{M}_\mathrm{bulge})$ \\
\hline
0.65 & 0.00601 & 0.00248 & 1.56 \\
0.95 & 0.00448 & 0.00110 & 2.02 \\
1.25 & 0.00373 & 0.00074 & 2.24 \\
1.60 & 0.00417 & 0.00095 & 2.09 \\
2.00 & 0.00292 & 0.00054 & 2.32 \\
2.40 & 0.00198 & 0.00051 & 1.97 \\
2.80 & 0.00169 & 0.00028 & 2.45 \\
\hline
\end{tabular}
\caption{\label{amplitude} \rev{Estimated change in the GWB characteristic strain amplitude when SMBH masses are predicted from either \msigma or \mmb. We find that SMBH masses inferred from velocity dispersion lead to an estimated increase in the amplitude by a factor of 2.1 on average across all redshifts considered here.}}
\end{center}
\end{table}

\rev{From the 15 year results of NANOGrav's PTA, the offset between the signal amplitude and the highest value predictions for the GWB amplitude is at least a factor of 2 though potentially more \citep{Agazie_2023, Antoniadis_2023, Reardon_2023, Arzoumanian_2020, Shannon_2015, McWilliams_2014, Middleton_2021, Zhu_2019, Bonetti_2018}.} An in-depth analysis of how our results affect predictions for the GWB will be presented in future work, but the initial estimate we provide here suggests an origin for this difference. It is uncertain at this point whether velocity dispersion or stellar mass is necessarily a better SMBH mass indicator. It is clear, however, that further investigation is necessary so that we can further understand why these relations differ so greatly.

Future work investigating our findings is necessary. A good test the MFP would involve obtaining velocity dispersion measurements for a sub-sample of the galaxies in this survey for $z > 1$, with even a relatively small sample it would be possible to quantify the accuracy of the MFP at $z > 1$. Measured velocity dispersion estimates are the first step for evaluating the potential evolution of the MFP, but to thoroughly analyze how SMBH mass scaling relations may change with time, dynamical mass estimates at $z > 1$ are needed. 30-m class telescopes, suitable for high-redshift observations, make this feat a realistic goal and will expand our understanding of how galaxies and their SMBHs evolve \citep{Gultekin_2019}. Aside from tests of the results we show here, extending our work to include a robust analysis of lower mass (M$_\mathrm{BH} < 10^8 \mathrm{M}_\odot$) black holes will inform our predictions for the Laser Interferometer Space Antenna (LISA) mission which will be vital in our characterization of black hole see formation. With upcoming missions and the continued refinement in GWB detection efforts, a full picture of the potential evolution of galaxy SMBH scaling relations can emerge.

\section{Summary}\label{summary}

In this paper we examined the difference between SMBH mass predictions when assuming \msigma\ versus \mmb. To do this we used the three-parameter relationship between galaxy stellar mass, effective radius, and velocity dispersion to infer velocity dispersion for galaxies up to $z = 3$. We created SMBH mass density functions for all galaxies in our sample for  $0.5 < z < 3$ and compared how using stellar mass versus MFP velocity dispersion affected inferred SMBH demographics. We found that the number of SMBHs with masses M$_\mathrm{BH} < 10^9\ \mathrm{M}_\odot$ was different between these relations, especially for $z > 1$. In particular we find that \msigma\ predicts a greater number of these high mass SMBHs. Our results suggest that the relationship between SMBH mass and stellar mass and/or velocity dispersion must evolve at high redshift. Assuming the local relations to be constant across time leads to substantial differences when extrapolated beyond $z = 0.5$, and this difference must be reconciled.

Our results do not inform us of the accuracy of either relation. It remains unclear whether one or both relations are evolving. Recent work has found that the stellar mass to SMBH mass relation may have evolved at least since $z \sim 2$ \citep{Zhang_2023}, but no evolution has been investigated for velocity dispersion. Circumstantial evidence from, e.g., red nugget galaxies, points toward \msigma\ being a more accurate predictor of SMBH mass at these higher redshifts \citep{Y_ld_r_m_2015, Ferre_Mateu_2015, Ferre_Mateu_2017}. Prediction and interpretation of the GWB from PTAs relies heavily on the assumptions made for the SMBH demographics at high redshift. Here we have shown that the choice in scaling relation used to infer high redshift SMBH mass can lead to meaningfully different demographics. If we are to refine our ability to explore the physics of galaxy and SMBH evolution at $z > 1$ we must also re-examine how the local scaling relations may evolve.

\section*{Acknowledgements}

The authors would like to thank Eric Bell and Rachel Bezanson for their helpful conversations. We additionally thank Anna de Graaff, Joel Leja, and Arjen van der Wel for readily sharing their knowledge and data with us.

CM acknowledges financial support through the University of Michigan’s Rackham Merit Fellowship Program. 
JS is supported by an NSF Astronomy and Astrophysics Postdoctoral Fellowship under award AST-2202388.

We thank the anonymous referee for their insightful comments.

Anishinaabeg gaa bi dinokiiwaad temigad manda Michigan Kichi Kinoomaagegamig. Mdaaswi nshwaaswaak shi mdaaswi shi niizhawaaswi gii-sababoonagak, Ojibweg, Odawaag, minwaa Bodwe’aadamiig wiiba gii-miigwenaa’aa maamoonjiniibina Kichi Kinoomaagegamigoong wi pii-gaa aanjibiigaadeg Kichi-Naakonigewinning, debendang manda aki, mampii Niisaajiwan, gewiinwaa niijaansiwaan ji kinoomaagaazinid.  Daapanaming ninda kidwinan, megwaa minwaa gaa bi aankoosejig zhinda akiing minwaa gii-miigwewaad Kichi-Kinoomaagegamigoong aanji-daapinanigaade minwaa mshkowenjigaade.

The University of Michigan is located on the traditional territory of the Anishinaabe people. In 1817, the Ojibwe, Odawa, and Bodewadami Nations made the largest single land transfer to the University of Michigan.  This was offered ceremonially as a gift through the Treaty at the Foot of the Rapids so that their children could be educated. Through these words of acknowledgment, their contemporary and ancestral ties to the land and their contributions to the University are renewed and reaffirmed.

\section*{Data Availability}


The data generated through this project will be deposited into Deep Blue Data, the University of Michigan's institutional data repository.  Data that we supply but is based on formatted versions of others' work will include attribution and notices that they are downstream products of others' work.


\bibliographystyle{mnras}
\bibliography{ref}


\appendix
\section{Fit Parameters}

The posterior fit parameters for stellar mass, velocity dispersion, and black hole mass functions are presented in the tables \ref{SMF}, \ref{VDF}, \ref{BHMFMM}, and \ref{BHMFmsig} found here. The errors listed are 68\% confidence intervals. Because of degeneracy between some of the fit parameters, e.g., $\phi_*$ and $\alpha$, the errors reported here are the confidence intervals on a given variable and are not the same as the 68\% confidence fits shown by the darker shaded region in each plot.

\begin{table*}
\begin{center}
\begin{tabular}{|ccccccc|}
\hline
Stellar Mass Posteriors \\
\hline
Median Redshift & $\log{\phi_{*, 1}}$ & $\log{\phi_{*, 2}}$ & $\alpha_{s, 1}$ & $\alpha_{s, 2}$ & $\log \mathrm{M}_\mathrm{c}$ & $\sigma_{\mathrm{scatter}}$ \\
\hline
& & & All Galaxies & & & \\
0.65 &$ -2.537 \pm 0.546 $ & $ -3.445 \pm 0.402 $ & $ -0.699 \pm 0.236 $ & $ -1.649 \pm 0.207 $ & $ 10.986 \pm 0.093 $ & $ -1.400 \pm 0.139 $ \\
0.95 &$ -4.548 \pm 0.532 $ & $ -2.850 \pm 0.040 $ & $ -0.111 \pm 0.164 $ & $ -1.165 \pm 0.030 $ & $ 11.072 \pm 0.080 $ & $ -1.351 \pm 0.151 $ \\
1.25 &$ -3.200 \pm 0.698 $ & $ -3.051 \pm 0.121 $ & $ -0.185 \pm 0.278 $ & $ -1.177 \pm 0.126 $ & $ 10.946 \pm 0.128 $ & $ -1.263 \pm 0.143 $ \\
1.60 &$ -4.671 \pm 0.498 $ & $ -3.109 \pm 0.067 $ & $ -0.094 \pm 0.175 $ & $ -1.216 \pm 0.057 $ & $ 11.170 \pm 0.093 $ & $ -1.284 \pm 0.168 $ \\
2.00 &$ -4.608 \pm 0.568 $ & $ -3.403 \pm 0.080 $ & $ -0.128 \pm 0.186 $ & $ -1.296 \pm 0.096 $ & $ 11.154 \pm 0.131 $ & $ -1.122 \pm 0.155 $ \\
2.40 &$ -4.814 \pm 0.362 $ & $ -3.610 \pm 0.179 $ & $ -0.067 \pm 0.121 $ & $ -1.449 \pm 0.122 $ & $ 11.144 \pm 0.162 $ & $ -1.189 \pm 0.141 $ \\
2.80 &$ -4.167 \pm 0.580 $ & $ -3.911 \pm 0.288 $ & $ -0.321 \pm 0.212 $ & $ -1.335 \pm 0.098 $ & $ 11.039 \pm 0.173 $ & $ -1.124 \pm 0.167 $ \\
    \hline
\end{tabular}
\caption{\label{SMF} Posterior results for stellar mass fits.}
\end{center}
\end{table*}

\begin{table*}
\begin{center}
\begin{tabular}{|ccccc|}
\hline
Velocity Dispersion Posteriors \\
\hline
Median Redshift & $\log{\phi_*}$ & $\alpha_s$ & $\log \sigma_\mathrm{c}$ & 
$\sigma_{\mathrm{scatter}}$ \\
\hline
& & All Galaxies & & \\
0.65 & $ -3.691 \pm 0.227 $ & $ 3.646 \pm 0.327 $ & $  1.578 \pm 0.035 $ & $ -0.882 \pm 0.176 $ \\
0.95 & $ -3.161 \pm 0.206 $ & $ 2.508 \pm 0.366 $ & $  1.676 \pm 0.045 $ & $ -0.952 \pm 0.153 $ \\
1.25 & $ -2.982 \pm 0.133 $ & $ 1.812 \pm 0.283 $ & $  1.770 \pm 0.045 $ & $ -0.947 \pm 0.145 $ \\
1.60 & $ -3.091 \pm 0.139 $ & $ 1.595 \pm 0.273 $ & $  1.884 \pm 0.050 $ & $ -0.989 \pm 0.166 $ \\
2.00 & $ -3.224 \pm 0.142 $ & $ 1.431 \pm 0.263 $ & $  1.922 \pm 0.055 $ & $ -0.943 \pm 0.162 $ \\
2.40 & $ -3.423 \pm 0.134 $ & $ 1.319 \pm 0.216 $ & $  1.963 \pm 0.065 $ & $ -0.894 \pm 0.168 $ \\
2.80 & $ -3.429 \pm 0.094 $ & $ 1.207 \pm 0.190 $ & $  1.944 \pm 0.061 $ & $ -0.879 \pm 0.161 $ \\
\hline
\end{tabular}
\caption{\label{VDF} Posterior results for velocity dispersion fits.}
\end{center}
\end{table*}

\begin{table*}
\begin{center}
\begin{tabular}{|ccccc|}
\hline
M$_\text{BH}$--$\text{M}_\mathrm{bulge}$ Posteriors \\
\hline
Median Redshift & $\log{\phi_*}$ & $\alpha_s$ & $\log \mathrm{M}_\mathrm{c}$ & 
$\sigma_{\mathrm{scatter}}$ \\
\hline
& & All Galaxies & & \\
0.65 & $ -3.180 \pm 0.090 $ & $ -1.262 \pm 0.028 $ & $  9.140 \pm 0.113 $ & $ -1.128 \pm 0.113 $ \\
0.95 & $ -3.312 \pm 0.106 $ & $ -1.250 \pm 0.040 $ & $  9.014 \pm 0.121 $ & $ -1.063 \pm 0.102 $ \\
1.25 & $ -3.433 \pm 0.097 $ & $ -1.210 \pm 0.039 $ & $  8.983 \pm 0.116 $ & $ -1.117 \pm 0.096 $ \\
1.60 & $ -3.659 \pm 0.123 $ & $ -1.312 \pm 0.047 $ & $  9.182 \pm 0.138 $ & $ -1.111 \pm 0.105 $ \\
2.00 & $ -3.889 \pm 0.194 $ & $ -1.309 \pm 0.065 $ & $  9.172 \pm 0.248 $ & $ -1.006 \pm 0.094 $ \\
2.40 & $ -4.968 \pm 0.451 $ & $ -1.575 \pm 0.093 $ & $  10.038 \pm 0.594 $ & $ -1.016 \pm 0.106 $ \\
2.80 & $ -4.509 \pm 0.457 $ & $ -1.446 \pm 0.176 $ & $  9.403 \pm 0.595 $ & $ -0.947 \pm 0.116 $ \\
\hline
& & Quiescent & & \\
0.65 & $ -3.336 \pm 0.242 $ & $ -1.135 \pm 0.143 $ & $  9.256 \pm 0.320 $ & $ -0.980 \pm 0.103 $ \\
0.95 & $ -3.798 \pm 0.348 $ & $ -1.208 \pm 0.194 $ & $  9.390 \pm 0.489 $ & $ -0.945 \pm 0.106 $ \\
1.25 & $ -3.573 \pm 0.264 $ & $ -0.843 \pm 0.252 $ & $  9.021 \pm 0.451 $ & $ -0.919 \pm 0.111 $ \\
1.60 & $ -3.640 \pm 0.153 $ & $ -0.824 \pm 0.162 $ & $  9.085 \pm 0.212 $ & $ -0.926 \pm 0.113 $ \\
2.00 & $ -4.078 \pm 0.263 $ & $ -0.924 \pm 0.219 $ & $  9.377 \pm 0.565 $ & $ -0.842 \pm 0.116 $ \\
2.40 & $ -4.077 \pm 0.285 $ & $ -0.823 \pm 0.314 $ & $  9.527 \pm 0.649 $ & $ -0.857 \pm 0.121 $ \\
2.80 & $ -4.907 \pm 0.381 $ & $ -1.026 \pm 0.678 $ & $  9.573 \pm 0.512 $ & $ -1.094 \pm 0.118 $ \\
\hline
& & Star-Forming & & \\
0.65 & $ -4.235 \pm 0.682 $ & $ -1.764 \pm 0.269 $ & $  9.424 \pm 0.600 $ & $ -1.009 \pm 0.13 $ \\
0.95 & $ -4.292 \pm 0.643 $ & $ -1.723 \pm 0.279 $ & $  9.338 \pm 0.570 $ & $ -1.039 \pm 0.162 $ \\
1.25 & $ -4.524 \pm 0.591 $ & $ -1.691 \pm 0.227 $ & $  9.458 \pm 0.585 $ & $ -1.053 \pm 0.138 $ \\
1.60 & $ -4.667 \pm 0.572 $ & $ -1.743 \pm 0.187 $ & $  9.576 \pm 0.565 $ & $ -1.072 \pm 0.144 $ \\
2.00 & $ -4.645 \pm 0.600 $ & $ -1.595 \pm 0.272 $ & $  9.552 \pm 0.636 $ & $ -0.983 \pm 0.146 $ \\
2.40 & $ -5.055 \pm 0.469 $ & $ -1.743 \pm 0.165 $ & $  9.525 \pm 0.460 $ & $ -0.991 \pm 0.148 $ \\
2.80 & $ -4.678 \pm 0.511 $ & $ -1.518 \pm 0.659 $ & $  9.331 \pm 0.648 $ & $ -0.918 \pm 0.173 $ \\
\hline
\end{tabular}
\caption{\label{BHMFMM} Posterior results for SMBH mass fits using \mmb.}
\end{center}
\end{table*}

\begin{table*}
\begin{center}
\begin{tabular}{|ccccc|}
\hline
\msigma\ Posteriors  \\
\hline
Median Redshift & $\log{\phi_*}$ & $\alpha_s$ & $\log \mathrm{M}_\mathrm{c}$ & 
$\sigma_{\mathrm{scatter}}$ \\
\hline
& & All Galaxies & & \\
0.65 & $ -3.502 \pm 0.136 $ & $ -1.209 \pm 0.032 $ & $  9.775 \pm 0.250 $ & $ -1.021 \pm 0.082 $ \\
0.95 & $ -3.827 \pm 0.151 $ & $ -1.244 \pm 0.029 $ & $  10.031 \pm 0.297 $ & $ -1.015 \pm 0.086 $ \\
1.25 & $ -3.987 \pm 0.146 $ & $ -1.225 \pm 0.029 $ & $  10.181 \pm 0.307 $ & $ -1.008 \pm 0.082 $ \\
1.60 & $ -4.156 \pm 0.172 $ & $ -1.267 \pm 0.031 $ & $  10.528 \pm 0.343 $ & $ -1.017 \pm 0.077 $ \\
2.00 & $ -4.287 \pm 0.155 $ & $ -1.215 \pm 0.028 $ & $  10.571 \pm 0.389 $ & $ -0.983 \pm 0.075 $ \\
2.40 & $ -4.780 \pm 0.185 $ & $ -1.284 \pm 0.039 $ & $  11.154 \pm 0.394 $ & $ -0.933 \pm 0.084 $ \\
2.80 & $ -4.573 \pm 0.191 $ & $ -1.170 \pm 0.055 $ & $  10.795 \pm 0.471 $ & $ -0.896 \pm 0.073 $ \\
\hline
& & Quiescent & & \\
0.65 & $ -3.951 \pm 0.283 $ & $ -1.239 \pm 0.099 $ & $  10.333 \pm 0.509 $ & $ -0.905 \pm 0.081 $ \\
0.95 & $ -4.131 \pm 0.230 $ & $ -1.156 \pm 0.087 $ & $  10.370 \pm 0.489 $ & $ -0.871 \pm 0.081 $ \\
1.25 & $ -4.139 \pm 0.201 $ & $ -1.059 \pm 0.102 $ & $  10.414 \pm 0.433 $ & $ -0.834 \pm 0.078 $ \\
1.60 & $ -4.146 \pm 0.158 $ & $ -1.019 \pm 0.082 $ & $  10.458 \pm 0.384 $ & $ -0.877 \pm 0.084 $ \\
2.00 & $ -4.331 \pm 0.183 $ & $ -0.962 \pm 0.100 $ & $  10.841 \pm 0.455 $ & $ -0.824 \pm 0.090 $ \\
2.40 & $ -4.418 \pm 0.225 $ & $ -0.959 \pm 0.149 $ & $  10.735 \pm 0.470 $ & $ -0.843 \pm 0.178 $ \\
2.80 & $ -5.074 \pm 0.314 $ & $ -0.974 \pm 0.339 $ & $  10.546 \pm 0.592 $ & $ -1.109 \pm 0.164 $ \\
\hline
& & Star-Forming & & \\
0.65 & $ -4.513 \pm 0.379 $ & $ -1.460 \pm 0.101 $ & $  10.607 \pm 0.571 $ & $ -0.944 \pm 0.100 $ \\
0.95 & $ -4.685 \pm 0.378 $ & $ -1.456 \pm 0.091 $ & $  10.755 \pm 0.546 $ & $ -0.964 \pm 0.094 $ \\
1.25 & $ -4.652 \pm 0.310 $ & $ -1.393 \pm 0.077 $ & $  10.747 \pm 0.508 $ & $ -0.955 \pm 0.095 $ \\
1.60 & $ -4.722 \pm 0.328 $ & $ -1.421 \pm 0.071 $ & $  10.734 \pm 0.529 $ & $ -0.979 \pm 0.099 $ \\
2.00 & $ -4.743 \pm 0.291 $ & $ -1.337 \pm 0.072 $ & $  10.911 \pm 0.553 $ & $ -0.955 \pm 0.099 $ \\
2.40 & $ -4.981 \pm 0.225 $ & $ -1.361 \pm 0.065 $ & $  11.019 \pm 0.414 $ & $ -0.901 \pm 0.094 $ \\
2.80 & $ -4.654 \pm 0.272 $ & $ -1.221 \pm 0.112 $ & $  10.750 \pm 0.511 $ & $ -0.873 \pm 0.104 $ \\
\hline
\end{tabular}
\caption{\label{BHMFmsig} Posterior results for SMBH mass fits using \msigma.}
\end{center}
\end{table*}

\bsp	
\label{lastpage}
\end{document}